\documentclass[preprint]{aastex}    
\input epsf

\newcommand{\ofour}{HDE\,269698}
\newcommand{\osix}{HDE\,270952}
\newcommand{\oseven}{AzV\,232}
\newcommand{\onine}{Sk\,$-66{\arcdeg}169$}

\newcommand{\iue}{\it IUE\/}
\newcommand{\hst}{\it HST\/}

\newcommand{\vlt}{VLT}
\newcommand{\fuse}{\it FUSE\/}

\newcommand{\vinf}{$v_\infty$}

\slugcomment{Submitted to Ap.J}
\shorttitle{Revised Temperatures for O Supergiants}
\shortauthors{Crowther et al.}
\begin{document}
\title{Revised Stellar Temperatures for Magellanic Cloud O Supergiants 
from {\fuse} and VLT-UVES Spectroscopy\footnote{
Based on observations made with the
NASA-CNES-CSA Far Ultraviolet Spectroscopic Explorer. FUSE is operated
for NASA by The Johns Hopkins University under NASA contract NAS5-32985.
Also based in part on observations collected at the European Southern 
Observatory Very Large Telescopes in programs 65.H-0705 and 
67.D-0238, plus archival data
obtained with the NASA-ESA Hubble Space Telescope and NASA-ESA-PPARC
International Ultraviolet Explorer.}}

\author{ P. A. Crowther\altaffilmark{2},
         D. J. Hillier\altaffilmark{3},
         C. J. Evans\altaffilmark{2},
         A. W. Fullerton\altaffilmark{4,5},
         O. De Marco\altaffilmark{2,6},
         A.~J.~Willis\altaffilmark{2}}


\altaffiltext{2}{Dept. of Physics \& Astronomy,
                 University College London,
                 Gower Street, London WC1E~6BT,
                 England}
\altaffiltext{3}{Department of Physics \& Astronomy, 
                 University of Pittsburgh, 3941 O'Hara Street, PA 15260}
\altaffiltext{4}{Dept. of Physics \& Astronomy, 
                 University of Victoria,
                 P.O. Box 3055, 
                 Victoria, BC, V8W 3P6, 
                 Canada}
\altaffiltext{5}{Center for Astrophysical Sciences,
                 Dept. of Physics \& Astronomy, 
                 The Johns Hopkins University,
                 3400 N. Charles Street, 
                 Baltimore, MD 21286}
\altaffiltext{6}{Department of Astrophysics,
                 American Museum of Natural History,
                 Central Park West at 79th St, New York, NY 10024}

\begin{abstract}
We have undertaken quantitative analysis of four
LMC and SMC O4--9.7 extreme supergiants
using far-ultraviolet {\fuse}, ultraviolet {\iue}/{\hst} 
and optical {\vlt} UVES
spectroscopy. Extended, non-LTE model 
atmospheres that allow for the consistent treatment of line blanketing 
(Hillier \& Miller 1998) 
are used to analyse wind and photospheric spectral features simultaneously.
Using H$\alpha$ to constrain $\dot{M}$, He\,{\sc i-ii} photospheric lines 
reveal stellar temperatures which are systematically 
(5--7.5kK) and substantially (15--20\%)
lower than previously derived 
from unblanketed, plane-parallel, non-LTE photospheric  studies. We
have confidence in these revisions, since derived temperatures 
generally yield consistent fits across the entire 
$\lambda\lambda$912--7000\AA\ observed spectral range.
In particular, we are able to resolve the UV-optical temperature discrepancy
identified for \oseven\ (O7\,Iaf$^+$) in the SMC by Fullerton et al. (2000).
%
  
The temperature and abundance sensitivity of far-UV, UV and optical lines 
is discussed. `Of' classification criteria are directly linked to (strong) 
nitrogen enrichment (via N\,{\sc iii} $\lambda$4097) 
and (weak) carbon depletion (via C\,{\sc iii} $\lambda\lambda$4647-51),
providing evidence for mixing of unprocessed and CNO processed 
material at their stellar surfaces. 
Oxygen abundances are more difficult 
to constrain, except via O\,{\sc ii} lines in 
the O9.7 supergiant for which it is also found to be somewhat depleted.
Unfortunately, He/H is very difficult to determine
in individual O supergiants, due to uncertainties in microturbulence 
and the atmospheric scale height. The effect of wind clumping 
is also investigated, for which P\,{\sc v} $\lambda\lambda$1118--28 
potentially provides a 
useful diagnostic in O-star winds, unless phosphorus can be independently
demonstrated to be 
underabundant relative to other heavy elements. Revised stellar properties
affect existing calibrations of (i)
Lyman continuum photons -- a factor of two lower for the O4
supergiant; and (ii) kinetic energy released into the ISM by O supergiants.
Our results also have importance for the calibration of the
wind momentum-luminosity relationship for OB stars, particularly since
the stars studied here are amongst the visually 
brightest OB stars in external galaxies.
\end{abstract}


\section{Introduction}

The existence of winds in O-type stars has been established 
since the 1960's, when the first rocket-ultraviolet (UV) observations 
%
revealed the characteristic resonance line P~Cygni signatures of 
mass loss (Morton 1967). Far-ultraviolet (FUV) spectroscopy of 
O-type stars with {\it Copernicus}, the {\it Hopkins Ultraviolet 
Telescope} and {\it ORFEUS} missions revealed many additional
stellar-wind features.
The launch of the {\it Far Ultraviolet Spectroscopic 
Explorer} ({\fuse}) telescope (Moos et al. 2000) has provided a new
opportunity to study a wide variety of OB stars spanning a range 
of metallicities, $Z$, at high spectral-resolution.

Of primary interest is the dependence of mass-loss rates for
luminous OB stars on $Z$. Theoretically, the strength of 
%
radiatively driven O-star winds is predicted to depend on metallicity,
$Z$, as
$\dot{M} \propto Z^{0.5-0.7}$ (Kudritzki, Pauldrach \& Puls 
1987; Vink, de Koter \& Lamers 2001). 
Observationally, the principal method of deriving mass-loss properties of 
O stars has been via radio observations for stars within a few kpc, 
or H$\alpha$ observations more generally (Puls et al. 1996). {\fuse} 
revives the possibility of using UV  resonance lines to determine 
mass-loss rates empirically. 

Observational results from H$\alpha$ have 
been combined with theoretical predictions to generate the so-called 
Wind-Momentum-Luminosity Relationship (WLR; Kudritzki \& Puls 2000), which
can be used to determine extragalactic distances. To date, such studies almost
exclusively employed plane-parallel, non-LTE models for determinations of
temperature (using photospheric He lines), and separately use wind models 
to determine mass-loss rates (using H$\alpha$ or UV wind profiles). 
Recent examples of the latter approach include Pauldrach et al. (1994)
and Pauldrach, Hoffman \& Lennon (2001). Clearly, the underlying assumption
in such studies is that the effects of winds on photospheric optical
lines is negligible -- see, however, 
Gabler et al. (1989) and Schaerer \& Schmutz (1994). Generally either
line blanketed or spherically extended model atmospheres have been employed
(e.g. Herrero, Puls \& Villamariz 2000). Rarely have both effects been
considered simultaneously.

Since OB supergiants have an enormous impact on the chemical and
dynamical evolution of their environments, their properties are of
considerable interest.
Recently,
Fullerton et al. (2000; hereafter Paper~I) presented a study of {\oseven}
(O7\,Iaf$^{+}$, SMC) and Sk-67$^{\circ}$ 111 
(O6\,Ia(n)fp var\footnote{Classification revised recently by 
Walborn et al. 2002a}, LMC)
based on {\fuse} spectroscopy. Non-LTE wind models allowing for 
line blanketing and an expanding atmosphere (Pauldrach et al. 2001) 
were used to constrain their stellar temperatures to $T_{\rm eff}\sim$32kK based on 
{\fuse} FUV wind profiles. In contrast, recent non-LTE, plane-parallel, 
hydrostatic  studies of {\oseven} (e.g. Puls et al. 1996), 
derived
a substantially higher stellar temperature of $T_{\rm eff}\sim$38kK from 
optical photospheric lines. 
This represents an important discrepancy with respect to derived 
bolometric luminosities ($\propto T_{\rm eff}^{4}$), and so indirectly 
affects the calibration of the WLR. 

In this paper, we investigate the properties of a small sample of extreme
OB supergiants in the Magellanic Clouds to verify whether the supergiants
studied in Paper~I are typical, and if so, how these discrepancies may
be resolved. These stars were selected to cover a wide range of spectral
types, with the additional criteria that (i) they should have 
 low interstellar H$_2$
column densities to minimize contamination in the {\fuse} region; and (ii) 
they should have small projected rotational velocities.
We supplement the {\fuse} spectroscopy with {\iue} and 
{\hst} UV spectroscopy, together with ground-based optical observations in 
order to help resolve previously conflicting determinations of stellar temperature 
and mass-loss rate. To date, UV studies of OB stars generally adopt 
temperatures from optical plane-parallel analyses (e.g. Haser et al. 1998).
The sole exception was HD\,93129A 
(O2~If*, Walborn et al. 2002b)
for which Taresch et al. (1997) consistently analysed its optical, UV, 
and FUV spectrum.

The paper is structured as follows. Observations of the four program
O supergiants are presented in \S~\ref{obs}, followed by a 
determination of stellar parameters using standard plane-parallel
methods in \S~\ref{hydro}. Spherical, line-blanketed models
are utilised in \S~\ref{sect4}, revealing a substantial revision in
stellar properties. Abundances, including CNO elements are discussed in
\S~\ref{abundances}, whilst clumping in O supergiants is considered in
\S~\ref{clumping}. Finally, our conclusions are reached in 
\S~\ref{conclusions}.

\section{Observations}\label{obs}

Basic observational quantities for our targets are 
provided in Table~\ref{t1}. All supergiants have "Ia$^+$" spectral
types, although the meaning of the "+" differs between the sample, namely
Si\,{\sc iv} $\lambda\lambda$4088--4116 emission for the O4--6 supergiants
and He\,{\sc ii} $\lambda$4686 emission in the O9.7 star.
Note that {\onine} is not HDE\,269889 
(Walborn, priv. comm.), in conflict with the 
SIMBAD database entry for this star. The observed energy 
distribution of {\onine} supports V=11.56 mag 
(Fitzpatrick 1988), rather than V=12.56 mag (Isserstedt 
1975).

\subsection{Far-UV Spectroscopy}

Spectra of our program stars were obtained as part of the {\fuse}
Principal Investigator Team programs P117 (P.I.: J. Hutchings) and P103 
(P.I.: K. Sembach), plus Cycle~1
Guest Investigator program A133 (P.I.: A.W. Fullerton) between 1999 
December and
2000 September. These observations were made in time-tag mode through the
{30\arcsec $\times$ 30\arcsec} (LWRS) aperture.
Total exposure times were
5~ksec ({\onine}), 12~ksec ({\oseven}), 27~ksec ({\osix}) and 222~ksec
({\ofour}, this is the mean spectrum obtained during a monitoring
program).
As described by Moos et al. (2000) and Sahnow et al. (2000), FUSE
data consist of spectra with a 
resolving power
of $\sim$15,000 from
two Lithium Fluoride (LiF) channels, which cover
$\lambda\lambda$ 990--1187~{\AA}, and two Silicon Carbide (SiC) channels,
which cover $\lambda\lambda$ 905--1105~{\AA}.

Spectra from each channel were processed by the current version of the
standard calibration pipeline (CALFUSE 2.0.5), which corrects for drifts
and distortions in the readout electronics of the detectors, removes the
effects of thermally induced grating motions, subtracts a background image,
corrects for residual astigmatism in the spectrograph optics, and applies
flux and wavelength calibrations to the extracted spectra.
Spectra from the individual channels were subsequently aligned, merged, and
resampled to a constant wavelength step of 0.13~{\AA} in the manner
described 
in
Walborn et al. (2002a).

Our calibrated spectra are shown in Fig.~\ref{fusedata},
along with identifications for important stellar features.
{\fuse} complements existing UV spectroscopy of OB stars by providing
high and low excitation
 P~Cygni resonance profiles, including important lines belonging to
C\,{\sc iii}, N\,{\sc iii}, O\,{\sc vi}, P\,{\sc v}, S\,{\sc iv, vi}. In 
\S~\ref{sect4}
we shall demonstrate the superior sensitivity of these diagnostics to
stellar temperature and abundances, relative to 
the usual UV P~Cygni resonance lines of C\,{\sc iv}, N\,{\sc v} and to
a lesser degree
Si\,{\sc iv}. Systematic trends in these features have been discussed in
detail by 
Walborn et al.
(2002a).

\subsection{UV Spectroscopy}

Complementary high-dispersion UV spectroscopy 
has
been obtained from 
the {\iue} and {\hst} data archives. All our program stars have been observed
with the {\iue} satellite, 
with
the large aperture (LAP) in the 
short-wavelength (SWP) channel at high resolution (HIRES). In the case of
{\oseven}, multiple exposures were obtained. 


In addition, {\oseven} and {\ofour} were observed by {\hst}--FOS in
1995 January, at a higher S/N than {\iue}, albeit with reduced spectral
resolution (see Walborn et al. 1995a). We present {\iue} and {\hst}
UV spectroscopy of  our program stars in Fig.~\ref{iue-hst}. 
From the presence of strong P Cygni 
N\,{\sc v}   $\lambda\lambda$1238-42, 
Si\,{\sc iv} $\lambda\lambda$1393-1402, and
C\,{\sc iv}  $\lambda\lambda$1548-51
profiles, 
tt is clear
that each possesses a very powerful stellar wind.

\subsection{Optical Spectroscopy}

Ground-based optical spectroscopy of our program stars 
was
obtained with 
the European Southern Observatory (ESO) 8.1~m Very Large Telescope (VLT) 
in Paranal, Chile. Supplementary long-slit datasets were also taken with 
the 2.3~m Australian National University (ANU) telescope, 
or 3.9~m Anglo-Australian Telescope (AAT), both in Siding Spring, Australia. 

The VLT observations were obtained during 2001 September 27--28
with the UV-Visual Echelle Spectrograph (UVES) mounted on Kueyen
(UT2).  
The Dichroic \# 2 was used with a standard
blue setting for CCD \#2 ($\lambda$=437nm) 
providing continuous coverage between 
$\lambda\lambda$3731 and 4999,
recorded on a single 2$\times$4K EEV CCD (15$\times15 \mu$m pixels).
A non-standard red
setting for CCD \# 4 ($\lambda$=830nm) with an identical EEV CCD
covered $\lambda\lambda$6370--8313, plus a 2$\times$4K 
MIT/LL CCD (15$\times 15\mu$m pixels), 
covered $\lambda\lambda$8290--10252.
 A 1$''$ wide slit was used in variable seeing conditions
(0.8--2$''$), providing a 2 pixel spectral resolution of 0.09~\AA\ 
at H$\alpha$.

We have used the flux distributions obtained from long-slit, 
intermediate-dispersion 
spectra of our targets to help correct for the grating
blaze function in our echelle data, particularly in the vicinity of
lines with broad emission wings (e.g., H$\alpha$).

Observations covering H$\alpha$ and He\,{\sc ii} $\lambda$5412 
were obtained with 
the Faint Object Red Spectrometer 2 (FORS2) also mounted on Kueyen (UT2)
during poor seeing conditions in 2000 September. FORS2 was used in
long-slit mode with a 600R grism, 2048$\times$2048 pixel CCD and 0.7$''$ 
slit, providing spectral coverage of $\lambda\lambda$5330--7540 and
3 pixel resolution of 3.6~\AA. 
Subsequently, 
2.3 m ANU observations were taken with the Double Beam Spectrograph 
(DBS) on 2000 December 12--15.  
The 1200 line\,mm$^{-1}$ blue and red gratings were used with the 
corresponding arms of DBS
plus identical 1752$\times$532 SITE CCDs
(15$\times15 \mu$m pixels), covering a spectral range of 
3969--4967~\AA\ (blue) and $\lambda$5750--6710\AA\ (red). 
A 1.5$''$ wide slit was used in moderate seeing conditions 
($\sim$1.2$''$) to achieve a 2 pixel spectral resolution of 1.2~\AA.

Complementary blue 
high-dispersion spectroscopy of {\osix} and {\onine}
was collected at the AAT with the 
UCL echelle spectrometer (UCLES) during 1997 January. 
The 31 line\,mm$^{-1}$ grating, 
Tektronix 1024$\times$1024 pix CCD and 2$''$ slit provided complete blue 
spectral coverage of $\lambda\lambda$3874--5093 at a resolution of 0.15 \AA\
at H$\gamma$. 
%

All datasets were cleaned of cosmic rays, 
bias corrected, flat fielded, 
and optimally extracted in {\sc iraf} (v2.11\footnote{IRAF is
written and supported by the National Optical Astronomy Observatory (NOAO) in 
Tucson, AZ; http://iraf.noao.edu/}).  Subsequent
reductions (wavelength correction and merging of echelle orders)
were carried out with {\sc figaro} (Shortridge et al. 1999) 
and related packages.

Since the blue optical spectral morphology for most of our sample 
has
recently been discussed elsewhere (e.g. Fitzpatrick 1991; Walborn 1977;
Walborn et al. 1995a), in Fig.~\ref{ha-atlas}
we merely show rectified {\vlt}  spectroscopy of our targets in 
the vicinity of H$\alpha$, once again
indicating powerful stellar winds in all cases. 

\subsection{Stellar Wind Velocities}

For the present targets, we present 
measurements of $v_{\rm black}$ (Prinja, Barlow \& Howarth 1990)
in Table~\ref{t2} for several UV and FUV lines,
in some cases updated from Prinja \& Crowther (1998). 
Note that the shape of 
the absorption trough of the {\ion{C}{3}} $\lambda$977 resonance 
line is strongly affected by absorption from interstellar Ly~$\gamma$,
so this line should generally not be used as a  {\vinf} indicator. 

In order to derive terminal velocities we require a reliable measurement
of the stellar radial velocity, which is obtained from UVES 
datasets, via optical He photospheric lines. We generally adopt terminal
velocities from N\,{\sc iii} $\lambda$990, which provides good 
consistency with the usual Si\,{\sc iv} $\lambda\lambda$1394-1407 and
C\,{\sc iv} $\lambda\lambda$1548-51 diagnostics. Uniquely for {\onine}, we 
adopt the (higher) terminal velocity from Si\,{\sc iv} instead. 
In general, reasonable consistency is achieved relative to 
Sobolev with Exact Integration (SEI) line profile modeling 
(Haser 1995; Haser et al. 1998; Massa et al. 2002) as indicated in 
Table~\ref{t2}.

\section{Stellar Parameters Derived from Plane-Parallel Hydrostatic Models}\label{hydro}

Before we discuss results for our program stars allowing for the presence
of stellar winds,
we first follow the usual method of determining 
stellar temperatures,  surface gravities and helium abundances, namely via 
the standard plane-parallel, hydrostatic methods as employed by Herrero et al. 
(1992) and Smith \& Howarth (1994).

\subsection{Technique}

A large grid of hydrostatic, plane-parallel model atmospheres
calculated with {\sc tlusty} (Hubeny \& Lanz 1998) were used, 
involving $T_{\rm eff}$, $\log g$ and He/H, as follows.
Parameters are found by mapping the locus of models that reproduce the
measured equivalent widths in the ($T_{\rm eff}$, $\log g$) 
plane for each optical helium line. The He\,{\sc ii} 
lines are strongly sensitive to temperature for O stars and
$\lambda\lambda$4200, 4541
are used as the primary temperature diagnostics
(He\,{\sc ii} $\lambda$4686 is not considered due to the 
wind effects for this line). Following Smith \& Howarth (1998),
a microturbulence of $\xi$ = 15 km\,s$^{-1}$ is adopted for
this part of the analysis. Values in the range 10--20 km\,s$^{-1}$ are
considered for our subsequent analysis based on extended, 
line-blanketed
model atmospheres (\S~\ref{sect4}).

The determination of $\log g$ from Balmer line wings is challenging 
for stars with such strong winds, since H$\delta$ 
is strongly blended with N\,{\sc iii} $\lambda$4097 and Si\,{\sc iv} 
$\lambda\lambda$4088--4116, and 
H$\beta-\gamma$ show emission in their red 
wings. Consequently, the values derived here are from H$\epsilon$ and H8.  
The wings of the lines are compared with the grid of model spectra and a 
$\chi^2$ value is found for each model and then the locus of minima is 
mapped in the ($T_{\rm eff}, \log g$) plane. 
 
Prior to 
Balmer line fitting,
a measure of the macroscopic broadening of the
lines is required to convolve with the models before comparison with
observations.  As in Herrero et al. (1992) the projected rotational 
velocity $v \sin i$ is found using the He\,{\sc i} lines.  A model 
spectrum that reproduces the observed helium equivalent widths is 
interpolated from the grid, mapped onto the observations and then the 
broadening of the convolved model for which $\chi^2$ is a minimum is 
taken as $v \sin i$. 
 
Stellar parameters are selected from the ($T_{\rm eff}, \log g$) fit 
diagram (e.g.
Smith \& Howarth 1998).  Ideally, the single point where the 
loci 
intersect 
gives the parameters for the
model spectrum that best describes the star.  A model spectrum is then
calculated with those parameters.  Comparison of the model with the
observations permits a qualitative ``by-eye'' inspection of the line fits
and small changes of order $\Delta T_{\rm eff}$ = 1kK or 
$\Delta \log g$ = 0.1 are made if the overall quality of the fits is 
improved.  The strongest weight was given to $\lambda$4388 and 
$\lambda$4922 for He\,{\sc i} lines (not available for {\ofour}).  

In principle the helium abundance is determined by choosing the fit
diagram with the smallest intersection region in the ($T_{\rm eff}, \log g$) 
plane for differing values of He/H.  However, no obvious improvement 
over the fits at solar abundance was revealed.  Initial values at solar 
abundance were taken from the fit diagrams and then if a consistent model 
fit could not be obtained higher values were investigated.  
If the new model gave more a consistent fit that value was adopted.

\subsection{Hydrostatic Model Results}\label{hydrostatic}

As an example of the fit quality achieved, we present {\sc tlusty} model
fits to UVES optical observations of 
{\ofour} in Fig.\ref{hde269698_tlusty}. 
Clearly, He\,{\sc i-ii} (photospheric) absorption 
lines are well reproduced, in contrast with (wind) emission at 
He\,{\sc ii} $\lambda$4686. H$\alpha$ is totally dominated by
wind emission, whilst other members of the Balmer series, up 
to and including H$\epsilon$, also suffer from wind contamination.
Consequently, a firm determination of He/H is extremely difficult for
such extreme O supergiants. 
Surface gravities rely principally on the blue wings of 
Balmer-series members.
As discussed elsewhere (e.g. Puls et al. 1996), the neglect of wind
contamination systematically
underestimates the true surface gravity by $\sim$0.05--0.1 dex.

Spectroscopic results are presented in Table~\ref{t3} for our program
stars, including previous results by Puls et al. (1996) 
for {\ofour} and {\oseven}. Overall, previous hydrostatic results are
supported by our {\sc tlusty} studies, except that a somewhat higher 
$T_{\rm eff}$=31kK is derived for {\onine} than $T_{\rm eff}$=28kK
 obtained by Lennon et al. (1997).

\section{Stellar Parameters Derived with Extended, Line-Blanketed Models}\label{sect4}

Fig.~\ref{hde269698_tlusty}
illustrates the successes and failures
of plane-parallel techniques. Photospheric profiles in OB stars can be readily
matched allowing determination of physical parameters, yet in many non-dwarf
O stars, wind contamination 
prevents robust results. Such techniques, including those presented
above, generally neglect the 
effect of metal line blanketing (see, however, Hubeny et al. 1998). 
This has importance for the ionization 
structure. Therefore, we 
have additionally calculated spherically extended models which
explicitly allow for line blanketing, in order to better constrain the 
fundamental properties of OB supergiants.

\subsection{Modeling codes}\label{codes}

Several model atmosphere codes are now
 available for modeling the photospheres and stellar winds of early-type
stars without making the traditional core-halo approximation
 (see Crowther 1999 for a summary). 
We have carried out test calculations using {\sc isa-wind} 
(de Koter, Schmutz \& Lamers 1993; 
de Koter, Heap \& Hubeny 1997), {\sc cmfgen} (Hillier \& Miller 1998) 
and {\sc wm-basic} (Pauldrach et al. 2001),  for which a reasonable 
degree of consistency in the emergent FUV and UV 
spectra and wind ionization structure was obtained. 
Calculations carried out 
at our request for $\zeta$ Pup with 
the Potsdam code (Gr\"{a}fener et al. 2002)
also show remarkably good consistency.

For the present application, our requirements include the need to
study wind  and photospheric features simultaneously, 
and to consider line blanketing. 
We have therefore selected {\sc cmfgen},
given that: (i) {\sc wm-basic} does not yet properly account for 
Stark broadening in optical photospheric
lines, despite its extremely thorough (albeit approximate) treatment 
of line blanketing and shocks; (ii)
the Sobolev assumption which makes {\sc isa-wind} so computationally quick,
also hinders its usefulness for realistic photospheric
modeling 
of
OB-type stars.
Although the stellar photosphere and the highly supersonic wind are 
accurately parameterized by this code, the interface between these two 
regimes is poorly represented by the Sobolev approximation.

{\sc cmfgen} solves the equations of statistical equilibrium, radiative
transfer and radiative equilibrium, and incorporates line blanketing directly
through use of a super-level approach
(Hillier \& Miller 1998). The ions included in our calculations
are presented in Table~\ref{table0}. We construct two different model 
atoms:
one appropriate for early O supergiants, the other for late O supergiants.

The input atmospheric structure, connecting the spherically extended
hydrostatic layers to the $\beta$-law wind, is achieved via a parameterized
scale height, $h$. This is defined relative to the surface gravity of
the star in Hillier et al. (2002), via:
\[ h = 1.2 \times 10^{-3} \frac{(1+\gamma)}{\mu(1-\Gamma)}\frac{T_e}{g} 
R_{\odot},\]
where $\gamma$ is the mean number of electrons per ion, $\mu$ is
the mean ionic mass, $\Gamma$ is the ratio of radiation pressure
to $g$, and $T_e$ is the local electron temperature in Kelvin. We
initially adopt $h=0.005 R_{\ast}$, which is subsequently revised based on
fits to the wings of He\,{\sc i} and Balmer lines.
At high Rosseland optical depth, the parameterized form of our 
velocity law may be replaced with  the
equivalent plane-parallel hydrostatic structure obtained 
from {\sc tlusty} (Hillier et al. 2002). For the present application
this option is not utilized, since the program O supergiants all possess
extremely extended atmospheres, such that use of hydrostatic models, even
at depth is questionable. Comparisons between {\sc tlusty}
and {\sc cmfgen} have been undertaken in the case of very low mass-loss
rates by Hillier \& Lanz (2001) and were found to be fully consistent
in the case of negligible spherical extension. 

The formal solution of the 
radiative-transfer equation to obtain
the final emergent spectrum is computed separately, and includes standard Stark
broadening tables for H\,{\sc i}, He\,{\sc i-ii}. Except where noted,
these calculations assume a radially dependent `microturbulence' of 
the form used by Haser et al. (1998), with $\xi_{\rm min}$=10 
km\,s$^{-1}$ at the base of the wind and $\xi_{\rm max}$=100 km\,s$^{-1}$
at $v_{\infty}$. Haser et al. (1998) and 
Hillier et al. (2002) discuss the effect of varying $\xi$ 
in 
O-star models.

It has been established that high ionization stage resonance
lines, most notably O\,{\sc vi} $\lambda\lambda$1032--38 and
N\,{\sc v} $\lambda\lambda$1238--42,  
can only be reproduced in most O stars by considering X-rays  
(e.g. Pauldrach et al. 1994, 2001; Haser et al. 1998). 
X-rays are thought to
originate in O star winds via the intrinsic instability of 
radiatively driven winds (e.g. Owocki, Castor \& Rybicki 1988;
Feldmeier 1995), and 
affect high ions via Auger-ionization (Cassinelli \& Olson 1979).
 However,
MacFarlane et al. (1993) demonstrated for $\zeta$ Pup that it is 
{\it solely} these ion fractions throughout the wind 
that are sensitive to the X-ray flux. Consequently, 
the inclusion or exclusion of X-rays is generally
not relevant to the determination of fundamental stellar parameters
and abundances. 

Further, fitting O\,{\sc vi} and N\,{\sc v} requires several
parameters to be varied, including shock temperature, emissivity,
X-ray luminosity, all of which are {\it a priori} unknown, and
some of which are likely degenerate. Therefore, we have chosen against
attempts to {\it derive} shock parameters for the program stars, but
instead {\it adopt} a uniform set of X-ray properties for all final
model calculations.
We use a rather soft two component Raymond \&
Smith (1977) X-ray spectrum, 
with 3$\times 10^{6}$K and 5$\times 10^{6}$kK, respectively. We
scale their volume filling factors to  ensure 
$\log L_{\rm X}/L_{\rm bol} \sim -5.4$. This is towards the high end
of observed values for O stars 
(Chlebowski, Harnden \& Sciortino 1989), but as we shall
demonstrate below, it should be stressed that the bolometric luminosities of 
O supergiants previously derived may have been overestimated 
through inadequate temperature determinations. Two further assumptions are
that the filling factor of the 5 million Kelvin component is
fixed at a factor of two smaller than the 3 million Kelvin component. 




\subsection{Technique}

For individual stars, our approach is as follows. We adjust the stellar
temperature\footnote{Defined, as is usual for an extended atmosphere,
as
the effective temperature corresponding to the radius at a Rosseland
optical depth of 20.}
 and mass-loss rate of an individual model until the `photospheric'
He\,{\sc ii} $\lambda$4542 and He\,{\sc i} $\lambda$4471 lines are matched,
and simultaneously, we vary the total mass-loss rate until 
the {\it shape} of H$\alpha$ is also
reproduced. If necessary, we adjust the exponent of the $\beta$-law until
H$\alpha$ is better reproduced (see e.g. Fig. 15 in Prinja et al. 2001).
For the present sample 1 $\leq \beta \leq$ 2, with a typical accuracy of
$\pm$0.2. 
Fig.~\ref{mdot} illustrates this approach for \ofour, via a series of
identical models except for increasing mass-loss rate. Note 
that some features are very sensitive to mass-loss rate (H$\alpha$,
He\,{\sc ii}), some are modestly affected (He\,{\sc i} $\lambda$4471,
N\,{\sc iv} $\lambda$4058), whilst others are rather insensitive (N\,{\sc iii} 
$\lambda\lambda$4634--41).  

This figure also illustrates that, increasing the mass-loss 
rate, with all other parameters held fixed, 
the ratio of the {\it photospheric} 
O-star
classification lines, He\,{\sc ii} $\lambda$4542 and He\,{\sc i} 
$\lambda$4471, changes significantly. This implies  that an O supergiant 
with a {\it strong} stellar wind may possess a {\it lower} stellar 
temperature  than an O dwarf of identical spectral type. The 
usual assumption that stellar winds do not affect optical
photospheric lines is invalidated for such stars, as previously
suggested by, e.g., Schaerer \& Schmutz (1994). Further, line blanketing
affects the ionization balance of helium via backwarming, such that
the effect is two-fold relative to H-He plane-parallel, hydrostatic models.
Martins, Schaerer \& Hillier (2002)  demonstrate
the considerable effect of including line blanketing for O dwarfs.

Fortunately, the central absorption strength in our primary 
diagnostic lines is rather
insensitive to adopted atmospheric scale height, as discussed in detail
by Hillier et al. (2002).
The scale height 
is varied in the final stages of the analysis to ensure good consistency 
with observed wings of Balmer and He\,{\sc i} lines. 

The helium content is held fixed
at He/H=0.2 by number, since this is typical of other extreme
O supergiants (e.g. Crowther \& Bohannan 1997). 
The weak dependence of, e.g., He\,{\sc i} $\lambda$4471
and He\,{\sc ii} $\lambda$4542 
on
changes in He/H, and subtle dependence on
temperature, atmospheric
scale height and microturbulence (Smith \& Howarth 1998),
means that this ratio is exceptionally difficult to measure in practice.
 As we shall show, all program O supergiants are at least partially CNO
processed, so some level of He enrichment is expected. Fortunately, the 
exact 
He content
of individual stars is not crucial to the 
fundamental temperature scale of O supergiants which we are 
primarily concerned with
here. For 0.1$\leq$He/H$\leq$0.4, the potential uncertainty introduced
into mass-loss rate determinations is at most $\sim$30\%, 
unless clumping plays a role (see \S~\ref{clumping}). 
Other elemental abundances are initially fixed at 0.4$Z_{\odot}$ 
(LMC) or 0.2$Z_{\odot}$ (SMC) as determined from oxygen abundances in
these galaxies (Dufour 1984; Russell \& Dopita 1990).
Subsequently, CNO abundances are varied in
order to better match ultraviolet and optical metal lines. 
{\ofour} was one of two LMC O supergiants included in the study of
Haser et al. (1998), who obtained 0.5--0.8$Z_{\odot}$ from model
fits to {\hst}-FOS spectroscopy. Abundance changes in non-CNO elements
(e.g. S, Fe) would lead to somewhat improved fits, but this was not 
attempted since it was not central to the primary goal of this work. 
Our only  exception to this is for phosphorus, since this is of 
relevance to the question of clumping in O stars (see \S~\ref{clumping}).


At each step in the iterative process we 
ensure that the predicted flux distribution matches the de-reddened energy 
distribution. Interstellar reddening laws follow Seaton (1979), Howarth (1983)
and Bouchet et al. (1985) for the Galaxy, LMC and SMC, respectively. An average
Galactic contribution of E(B-V)=0.04  (SMC) or 0.07 mag (LMC) is assumed
following COBE/DIRBE results (Schlegel, Finkbeiner \& Davis 1998).
Distances of 50\,kpc and 60\,kpc are adopted 
for
the LMC and SMC, respectively (Westerlund 1997).

In all cases, we have selected targets with minimal 
molecular-hydrogen
column densities (Tumlinson et al. 2002). Therefore the 
strongest
interstellar contribution to the FUV spectrum is the atomic hydrogen
series, which has been taken into account following fits to Ly$\alpha$, 
following Herald, Hillier \& Schulte-Ladbeck (2001).

In order to avoid unnecessary repetition, we first discuss results for 
{\ofour} 
in detail, followed by
a concise summary for 
the remaining stars. The effect of $\dot{M}$ has already 
been presented for \ofour\ in Fig.~\ref{mdot}
so we limit our subsequent discussion to the effects
of temperature and abundance variations 
on
optical, UV and FUV spectral diagnostics.

\subsection{Multi-wavelength study of \ofour (O4\,Iaf$^+$)}

\subsubsection{Optical diagnostics}

In Fig.~\ref{hde269698_t} we compare optical observations of {\ofour}
with synthetic spectra for a wide range of temperatures, 34kK$\leq
T_{\rm eff} \leq $46kK,
selected to reproduce the observed H$\alpha$ emission and absolute  visual
magnitude, $M_{\rm V}$. 
For each temperature, two models are presented -- one in which
CNO elements are fixed at 0.4$Z_{\odot}$ (dotted), with the other N-rich
($\epsilon_{N}=6.0 \epsilon_{N,\odot}$)
and CO-poor ($\epsilon_{C}=0.05 \epsilon_{C,\odot}$, 
$\epsilon_{O}=0.1 \epsilon_{O,\odot}$) (solid). The latter
values are those which will be subsequently determined
for {\ofour} in the present section, whilst
X-rays are {\it neglected} for this part of the analysis, and
shall only be considered in the final model comparisons (see \S~\ref{o4uv}).
The parameters from these models, spanning 0.4\,dex in
luminosity, are listed in Table~\ref{t4}. The optical spectrum of
this star has previously been studied by Puls et al. (1996), whilst
Haser et al. (1998) analysed its {\sc HST}-FOS spectrum.

Fig.~\ref{hde269698_t} reveals that for the 
parameter space
investigated, some spectral features (N\,{\sc iii-v} and He\,{\sc i})
are very sensitive to $T_{\rm eff}$, 
whilst others (He\,{\sc ii}) are not. For lower temperature models 
of $T_{\rm eff}\sim$25kK, 
which we shall discuss below 
for {\onine}, He\,{\sc i-ii} sensitivities are reversed. 
From \S~\ref{hydrostatic}, hydrostatic fits to the optical  
He\,{\sc i-ii} lines of {\ofour} indicate $T_{\rm eff}\sim$46.5kK.
Our unified code reveals that negligible He\,{\sc i} $\lambda$4471 (the only
prominent blue optical He\,{\sc i} diagnostic) 
is predicted at such a high temperature. 
This line favours a much lower stellar temperature of $T_{\rm eff}$=
40$\pm$1 kK. 

Variable CNO abundances do not affect the strength of optical He\,{\sc
i-ii} lines, but some optical CNO lines are extremely abundance sensitive
(N\,{\sc iii} $\lambda\lambda$4634-41, C\,{\sc iii} $\lambda\lambda$4647-51;
N\,{\sc iv} $\lambda$4058).  Weak emission from S\,{\sc iv} 
$\lambda\lambda$4486-4504 is predicted at the lowest temperatures, supporting
the positive
identification of these features by Werner \& Rauch (2001).

For {\ofour} we are unable to
match the observed optical spectrum with cosmic CNO abundances -- the same is
true for all Of supergiants with strong N\,{\sc iii} $\lambda\lambda$4634--41 and
weak C\,{\sc iii} $\lambda\lambda$4647-51. For a N-rich, C-poor
atmosphere we are able to find a good match to the optical spectrum
at $T_{\rm eff}\sim$39 kK from fits to N\,{\sc iv}
$\lambda$4058, N\,{\sc v} $\lambda\lambda$4603--20, N\,{\sc iii}
$\lambda\lambda$4634--41 and N\,{\sc iv}
$\lambda\lambda$7103-29 (not shown). 
Oxygen diagnostics are absent from the optical, with the possible
exception of O\,{\sc iii} $\lambda$5592\footnote{O\,{\sc iii} $\lambda$5592
is not covered by our high quality UVES spectroscopy for any of our targets.},
such that we are unable to determine oxygen abundances. 
Significantly higher 
temperatures are excluded by the observed
weakness of N\,{\sc v} $\lambda\lambda$4603--20 and 
strength of $\lambda\lambda$4634--41. 
Taresch et al. (1997) used the N\,{\sc v}
$\lambda\lambda$4603--20 lines to constrain the temperature for HD\,93129A 
(O2\,If). Since `Of' stars are defined via 
N\,{\sc iii} $\lambda\lambda$4634--41 and He\,{\sc ii} 
$\lambda$4686 emission, one might argue that 
{\it all} Of supergiants exhibit partially
processed CNO surface abundances (see also the discussion in Voels et al. 1989). 

The broad emission feature in the vicinity of N\,{\sc iii} 
$\lambda\lambda$4634--41 and He\,{\sc ii} $\lambda$4686 is due to
incoherent electron scattering of line photons from these
multiplets (Hillier et al. 2002). Electron scattering 
also affects the Stark wings of the Balmer series, of
relevance for surface gravity determinations.

Since the stellar wind of {\ofour} is so powerful, wind and line 
blanketing 
dramatically affect the ionization balance of helium and nitrogen, 
such that a much lower stellar temperature is appropriate than deduced
from pure H-He hydrostatic models, resulting in a reduced luminosity, 
from 1.6$\times 10^{6}L_{\odot}$ for $T_{\rm eff}$=46.5kK in
the plane-parallel calculations to 9.5$\times 10^{5}L_{\odot}$
for $T_{\rm eff}$=40kK obtained here. Bohannan et al. (1986, 1990) have
previously emphasised the effect of wind blanketing for the temperature 
of $\zeta$ Pup (O4\,I(n)f), morphologically very similar to {\ofour},
implying a reduction from $T_{\rm 
eff}$=46.5kK to 42kK. The greater effect obtained here is due to 
line blanketing and the use of a spherical atmosphere instead of their
`core-halo' approach. Additionally, the 
inclusion of metal species allows us to use 
alternative temperature diagnostics, which give essentially identical
results\footnote{We admit a formal uncertainty of $\pm$1kK on our derived
stellar temperatures, so that the value obtained from optical
He\,{\sc i-ii} lines (40kK) is
indistinguishable from that using N\,{\sc iii-v} ($\sim$39kK)}.

For our N-rich model atmosphere, 
solely N\,{\sc iii} $\lambda$4379, not presented in our figures, 
is badly predicted (too strong in absorption) for 
$T_{\rm eff}=40$kK. We have investigated this 
failure, which cannot easily be attributed to incorrect N\,{\sc iii} 
atomic data, nor is it unique to 
{\ofour}, since other stars in our sample suffer from similar
deficiencies (see also Hillier et al. 2002). 
Further investigations are presently underway, but unlike 
$\lambda\lambda$4634--41 where a well known process (dielectronic recombination)
produces emission, no such mechanism is known for $\lambda$4379.
N\,{\sc iii} $\lambda\lambda$4510--4547 is not well reproduced either, but
this is directly attributable to our model atom, since few quartet
levels are considered, and further allowance needs to be made for
autoionization processes.

\subsubsection{UV diagnostics}\label{o4uv}

In contrast with most previous studies, which tend to concentrate
solely on optical {\it or} UV modeling, we additionally compare 
UV and FUV observations with our synthetic spectra, {\it determined}
from our optical analysis. In this way, one might be able to have 
confidence in results obtained using only optical temperature 
diagnostics, for those stars without UV observations.

Fig.~\ref{hde269698_t3} displays the rectified {\hst} spectroscopy of
{\ofour} covering $\lambda\lambda$1220--1750, together with synthetic spectra
from our models discussed above, again omitting X-rays for the moment.

 C\,{\sc iv}
$\lambda\lambda$1548--51, N\,{\sc iv} $\lambda$1718 
and He\,{\sc ii} $\lambda$1640 are remarkably
insensitive to stellar temperature in this 
temperature range.
In contrast, weaker wind features do respond to temperature.
Si\,{\sc iv} $\lambda\lambda$1393--1402 greatly increases in strength at
low temperatures, whilst  O\,{\sc v} $\lambda$1371 becomes prominent
at high temperatures and S\,{\sc v} $\lambda$1501 is present except
at the highest temperatures. O\,{\sc v} $\lambda$1371 has previously
been used as a temperature indicator for early O supergiants by
de Koter, Heap \&  Hubeny (1997), together with O\,{\sc iv} $\lambda\lambda$1338-43. In addition, 
the iron forest is particularly sensitive to stellar temperature. At low
temperatures Fe\,{\sc v} features between 
$\lambda\lambda$1350 and 1500 are very
prominent, with Fe\,{\sc iv} lines between 
$\lambda\lambda$1550 and 1700 present,
whilst
at high temperatures, Fe\,{\sc vi} is strong between
$\lambda\lambda$1250 and 1350,
with Fe\,{\sc v} weaker and Fe\,{\sc iv} absent. 
N\,{\sc v} $\lambda\lambda$1238--42
is rather temperature sensitive, but its strength is generally more 
indicative of X-rays than temperature. 

Turning to abundance diagnostics, in contrast with optical
CNO metal lines, UV lines are in general extremely poor indicators of
CNO abundance, with few exceptions. N\,{\sc iv}] $\lambda$1486, $\lambda$1718 
and O\,{\sc iv} $\lambda\lambda$1338-43 reveal (generally weak) abundance 
sensitivity.

Overall, 
intermediate-temperature models 
provide the best match to the iron forest, 
although the accuracy with which one can derive $T_{\rm eff}$ in this
way is lower than that offered by 
optical-line diagnostics.
optical line diagnostics. 
A high stellar temperature of 
$T_{\rm eff}\sim$46kK, as obtained in \S~\ref{hydrostatic}
from plane-parallel optical methods,
is firmly excluded by the observed weakness of Fe\,{\sc vi} and O\,{\sc v}
$\lambda$1371. Haser et al. (1998) encountered  problems with the latter
in their UV analysis of {\ofour} since they adopted $T{\rm eff}$=47.5kK
from Puls et al. (1996).
A lower temperature of $T_{\rm eff}\sim$35kK is actually 
favored
by the 
strength of Si\,{\sc iv}, although Fe\,{\sc iv} becomes too strong at this
temperature. N\,{\sc iv}] $\lambda$1486 is well matched with our N-rich 
atmosphere, whilst N\,{\sc iv} $\lambda$1718 is much better
reproduced with a normal N abundance. Overall, the UV spectrum of {\ofour} is 
broadly consistent with a N-rich, C-poor model with
$T_{\rm eff}$=37.5$\pm$2.5kK.

Finally, let us turn to the FUV. {\fuse} spectroscopy of {\ofour} between
$\lambda\lambda$910 and 1175
is presented in Fig.~\ref{hde269698_t2},
together with synthetic spectra from our set of models presented above. 
In contrast with the {\hst} spectral
region, the low- and high- ionization wind features now provide excellent
temperature diagnostics, whilst iron features are weak or absent.
The temperature sensitivity of 
C\,{\sc iii} $\lambda\lambda$977, 1175 and 
N\,{\sc iii} $\lambda\lambda$989--91 is
striking, as is the abundance sensitivity of
the latter. No match to $\lambda$989--91 can be achieved without N enrichment,
whilst P~Cygni emission at 
N\,{\sc iv} $\lambda$955 and $\lambda$923 is a little too strong at $T_{\rm eff}$=40kK, although the latter is heavily contaminated by the interstellar
H\,I Lyman series. 

The observed strength of $\lambda$1175 argues in favour of C depletion
for $T_{\rm eff}\sim$40kK.  As with Si\,{\sc iv} in the {\hst} spectrum, 
S\,{\sc iv} $\lambda\lambda$1062-1073 suggests a lower temperature of 
$T_{\rm eff}\sim$37kK. Its strength is affected by the adopted CNO 
abundance since these act as the principal wind coolants.
 P\,{\sc v} $\lambda\lambda$1118--28 (see below) and N\,{\sc iv} $\lambda$955
prove relatively insensitive. Unusually amongst the present sample, 
S\,{\sc vi} $\lambda\lambda$933-44 is
well reproduced in the (strong) blue P~Cygni component, but not in the 
(weak) red component, suggesting that this component is affected by a
feature absent from the models (Walborn et al. 2002a).   
P\,{\sc iv} $\lambda$950 is not
responsible since this is accounted for in our models.

Finally, it is apparent that the O\,{\sc vi} $\lambda\lambda$1032--38 
doublet is weakly present in {\ofour}. 
Solely the $T_{\rm eff}$=46kK synthetic spectrum reveals any 
prominent O\,{\sc vi} wind feature, albeit much weaker than observed. 
As discussed in \S~\ref{codes}, X-rays are generally
required in order to reproduce the strength of such `super-ions' in 
O stars (e.g. Pauldrach et al. 1994, 2001), which are omitted for the
moment. Our final {\ofour} fit does include X-rays, producing stellar
O\,{\sc vi} $\lambda\lambda$1032--38 with approximately the correct
strength.

In contrast with Paper~I, allowance for the stellar wind and line blanketing
permits broadly consistent results to be obtained for {\ofour} using optical,
UV and FUV diagnostics, as presented in  
Figs.~\ref{hde269698_cmfgen}-\ref{hde269698_cmfgen3}, which 
show our final fits to HDE\,269698 for
$T_{\rm eff}$=40kK,  $\log (L/L_{\odot})$=5.98,
     $\log g$=3.6, $\dot{M}$=8.5$\times 10^{-6}$ $M_{\odot}$yr$^{-1}$, 
$v \sin i$=80 km\,s$^{-1}$, 
     $\beta$=1 and $v_{\infty}$=1750 km\,s$^{-1}$,
as discussed above except that X-rays are now considered. 
With this set of parameters, overall consistency
is excellent with the exception of 
N\,{\sc iv} $\lambda\lambda$1718, 4058 
(too strong), Si\,{\sc iv} $\lambda\lambda$1393-1402 (too weak), 
S\,{\sc iv} $\lambda\lambda$1062-1072 (too weak) and P\,{\sc v} 
$\lambda\lambda$1118-28 (too strong -- see \S~\ref{clumping}). 

In contrast with the present results, which employ a single stellar atmospheric
code, previous studies of {\ofour} used a variety of tools. For example,
the photospheric optical lines were analysed using a conventional
plane-parallel hydrostatic non-LTE model by Puls et al. (1996), whilst
fits to H$\alpha$ were carried out using a different procedure. Haser
et al. (1998) subsequently used a {\it third} technique -- selecting a
 spherical, extended non-LTE model atmosphere to analyse the UV spectrum.

\subsection{Other O supergiants}

We shall now discuss the other three O supergiants studied in this work,
ordered by spectral type. Final stellar parameters for all our 
program stars are listed in Table~\ref{t5}, and compared with
previous results for these stars. In all previous cases temperatures
are derived from hydrostatic, plane-parallel models (e.g. Herrero
et al. 1992), whilst mass-loss
rates are obtained from H$\alpha$ modeling (e.g. Puls et al. 1996)
and wind velocities are obtained from SEI line profile modeling  (Haser
1995). For each star, 
optical and  UV/FUV synthetic spectra from these parameters are
compared with observations. Bolometric corrections (B.C.) differ
from the Vacca, Garmany \& Shull (1996) $T_{\rm eff}$--B.C. 
calibration by typically $-$0.1\,mag. Temperatures
are again revised downward, such that B.C.=$-3.2$ mag for {\oseven}
($T_{\rm eff}$=32kK) versus B.C.=$-$3.6 mag ($T_{\rm eff}$=37.5kK)
obtained previously by Puls et al. (1996).

Spectroscopic analysis of {\osix} and {\oseven} proceeds in a similar manner
to {\ofour}, 
with
an identical atomic model (Table~\ref{table0}). 
Optical He\,{\sc i-ii} diagnostics provide the 
best-fit temperature 
whilst H$\alpha$ provides the mass-loss rate in each case. As with {\ofour},
the only way in which we are able to reproduce the observed spectrum is
via N- enrichment and C- depletion in each case. Oxygen is again 
much more difficult to constrain. X-rays are included in the final
model atmosphere calculations, albeit with generic parameters.


\subsubsection{\osix (O6\,Iaf$^+$), LMC}

For {\osix},
T$_{\rm eff}$=33.5kK, $\dot{M}$=$1.1\times 10^{-5}$ $M_{\odot}$ yr$^{-1}$, 
$\log (L/L_{\odot}$)=5.86, $h=0.002 R_{\ast}$, and a supersonic velocity 
law of $\beta$=1.3 provides an overall 
excellent match to the optical spectrum as illustrated 
in Fig.~\ref{hde270952_cmfgen}. 
The $\beta$ law was principally determined from fits to H$\alpha$
and He\,{\sc ii} $\lambda$4686
(see Hillier et al. 2002 for the sensitivity of other 
O-star line profiles
to $\beta$). 
He\,{\sc ii} $\lambda\lambda$4200, 4542
absorption profiles are reproduced well, as is the absorption of
He\,{\sc i} $\lambda$4471, although there is a P~Cygni emission
in this feature which is not reproduced. The same is true for H$\beta$
and perhaps H$\gamma$. We predict an emission component of He\,{\sc i}
$\lambda$6678 in the observed blend with He\,{\sc ii} $\lambda$6683,
but the observed emission is stronger and sharper than predicted.

We derive 
$\epsilon_{\rm N}=4.5 \epsilon_{\rm N,\odot}$
for {\osix} using N\,{\sc iii} $\lambda$4097. A higher
abundance is suggested by $\lambda\lambda$4634-41, although other
N\,{\sc iii} features  (e.g. $\lambda\lambda$4510-47) 
argue for a lower N abundance. In contrast, the weak C\,{\sc iii}
$\lambda\lambda$4647--51
multiplet argues for depleted C abundance of
$\epsilon_{\rm C}=0.1 \epsilon_{\rm C,\odot}$. As with {\ofour},
the absence of any strong optical oxygen diagnostics prevent 
a reliable determination, so 
$\epsilon_{\rm O}=0.2 \epsilon_{\rm O,\odot}$ 
is adopted for oxygen.

In the UV, the {\iue} HIRES  SWP dataset is of rather low S/N. 
Nevertheless, the  Fe\,{\sc iv-v} forest, and  
Si\,{\sc iv} $\lambda\lambda$1393-1402 are well reproduced, except
for the strength of the red emission component, 
for T$_{\rm eff}$=33.5kK
as shown in Fig.~\ref{hde270952_cmfgen3}.
The superior diagnostics offered by the FUV {\fuse} region
also support the optically derived temperature and abundances.
S\,{\sc iv} $\lambda\lambda$1062--73, 
C\,{\sc iii} $\lambda\lambda$977, 1175 
and N\,{\sc iii} $\lambda\lambda$989--91 are all well matched,
as is the S\,{\sc vi} $\lambda\lambda$933-44 region, in contrast with
{\ofour}.  
As for {\ofour}, an improved fit to P\,{\sc v} $\lambda\lambda$1118--28
can be achieved via clumping (see \S~\ref{clumping}) 
or a reduced phosphorus abundance. N\,{\sc iv} $\lambda$955 
is rather too strong, suggesting a reduced N-abundance, although the 
(theoretically) more 
abundance-sensitive 
N\,{\sc iv}] $\lambda$1486 emission line
is matched rather well, as is N\,{\sc iii}] $\lambda$1750. 
The only prominent deficiency
in the UV/FUV synthetic spectrum is  N\,{\sc v} $\lambda\lambda$1238--42,
which is rather weak, despite the inclusion of X-rays, suggesting the need
for a harder X-ray spectrum and/or higher filling factors.


\subsubsection{\oseven (O7\,Iaf$^+$), SMC}

The optical spectral morphology of {\oseven} is rather 
similar to that of {\osix}
except for a somewhat weaker 
emission-line spectrum. 
Consequently, the
derived stellar parameters are  broadly similar. Hillier et al. (2002) 
describes in detail {\sc cmfgen} model comparisons for
AzV\,83 (also O7\,Iaf$^+$), a spectroscopic twin of {\oseven} (Walborn
et al. 2000) using  identical techniques, so we  defer to their more 
extensive discussion. For {\oseven},  we determine 
$T_{\rm  eff}$=32kK, $\dot{M}$=4.5$\times 10^{-6}$ $M_{\odot}$ yr$^{-1}$, 
$\log (L/L_{\odot}$)=5.85, $h$=0.005$R_{\ast}$ and a supersonic velocity 
law with $\beta$=1.65,
as presented in Fig.~\ref{av232_cmfgen}. These parameters
are in close agreement with those obtained from our analysis of
{\fuse} spectroscopy in Paper~I. The
revision in parameters for {\oseven} versus Puls et al. (1996) 
are as great as for {\ofour} (Table~\ref{t5}). 

From H$\alpha$ modeling we obtain a
slightly slower $\beta$-law than Puls et al. who obtained $\beta$=1.4,
also from H$\alpha$. In general, there is a tendancy for exponents 
determined from UV SEI line  profile modeling to be lower.
Haser (1995) obtained $\beta$=1.0 from N\,{\sc v}
$\lambda\lambda$1238--42, Si\,{\sc iv} $\lambda\lambda$1393--1402, 
C\,{\sc iv} $\lambda\lambda$1548--51, and N\,{\sc iv} $\lambda$1718 {\oseven},
although the quality of fits are of similar quality to that obtained 
for these profiles in the present  study. This difference should be taken
 into consideration when comparing UV 
and optically derived velocity laws. 
Puls et al. (1996) discuss similar discrepancies for $\zeta$ Pup.

Optical H and He line profiles
are well matched, with the exception of H$\beta$ (the central emission
is due to incorrectly subtracted nebular H$\beta$ from NGC\,346).
We obtain $\epsilon_{\rm N}=2.0 \epsilon_{\rm N,\odot}$ from the fit
to N\,{\sc iii} $\lambda$4097, whilst $\epsilon_{\rm C}=0.07 \epsilon_{\rm 
C,\odot}$ from 
C\,{\sc iii} $\lambda\lambda$4647-51.
Again, oxygen is poorly 
constrained from the optical, with $\epsilon_{\rm O}=0.1 \epsilon_{\rm 
O,\odot}$ adopted.  UV ({\hst}-FOS) and FUV ({\fuse})
comparisons are again very good, 
as illustrated in Fig.~\ref{av232_cmfgen3}, including C\,{\sc iii}, 
N\,{\sc iii}, S\,{\sc iv}, Si\,{\sc iv}, C\,{\sc iv} although N\,{\sc iv}
$\lambda$1718 is too strong, and He\,{\sc ii} $\lambda$1640 too weak.
This latter feature is reproduced much better when compared to higher 
resolution, albeit lower S/N {\iue} 
high-dispersion datasets.
The predicted P\,{\sc v}  $\lambda\lambda$1118-28 absorption is too strong, 
which as we shall show in \S~\ref{clumping} is either indicative of 
clumping or a reduced P abundance.  The main
failure for our {\oseven} model is N\,{\sc v} 
$\lambda\lambda$1238-42, again, for the adopted X-ray parameters.
To reiterate, the stellar parameters derived from our analysis are
 insensitive to these X-ray parameters.

\subsubsection{\onine (O9.7\,Ia$^+$), LMC}

For {\onine}, the lower observed ionization led to the selection of a 
slightly different atomic model (Table~\ref{table0}), whilst a
higher microturbulence of $\xi$=20 km\,s$^{-1}$ reproduced observed
optical diagnostic 
line profiles better than the generic $\xi$=10 km\,s$^{-1}$. 
Villamariz et al. (2002) recently derived $\xi$=20 km\,s$^{-1}$ in
their analysis of the Galactic O9.5\,Ib supergiant HD\,209975.
An optimum fit to He\,{\sc i} ($\lambda\lambda$4121, 4338, 4471, 4713) 
and He\,{\sc ii} ($\lambda\lambda$4200, 4542) absorption lines reveals
$T_{\rm eff}$=26\,kK, whilst H$\alpha$ indicates $\dot{M}$=
6.5$\times 10^{-6}$ $M_{\odot}$ yr$^{-1}$,
$\log (L/L_{\odot}$)=5.86, and $h=0.002R_{\ast}$, 
again 
with a velocity law of $\beta$=1.75 -- see
Fig.~\ref{sk66169_cmfgen}. This figure also shows that the 
Si\,{\sc iv}  $\lambda\lambda$4088-4116 and 
Si\,{\sc iii} $\lambda\lambda$4552-4576 lines 
are also consistently matched for this
temperature, although note the poor agreement with H$\delta$, 
uniquely for this star. Unusually, at least for the present sample, 
He\,{\sc ii} $\lambda$4686 emission is strongly overestimated for
our derived stellar temperature. Reproducing the strength of $\lambda$4686
in O stars is generally problematic, even using spherically
extended non-LTE models (e.g. Herrero et al. 2000). 
In order to reproduce its observed strength in {\onine}, we require a slightly
lower temperature of $T_{\rm eff}$=24kK. At this
lower temperature, the Pickering He\,{\sc ii} series, $\lambda$4542, 
$\lambda$4200 become very weak, although 
the He\,{\sc i} lines are essentially unchanged, with poorer agreement
for the Si\,{\sc iii-iv} diagnostics. Therefore, 
we adhere to our preferred solution of  $T_{\rm eff}$=26kK despite this 
obvious failure. Adopting a clumped model for {\onine} would help to
resolve this discrepancy, since He\,{\sc ii} $\lambda$4686 emission is
reduced, with negligible effect on the photospheric lines.

In addition to Si, CNO optical lines are generally well 
reproduced at  this temperature, requiring
$\epsilon_{\rm N}=2.0 \epsilon_{\rm N,\odot}$ from N\,{\sc iii} 
$\lambda$4097 and 
N\,{\sc ii} $\lambda\lambda$4601-43. 
Carbon and oxygen
abundances of $\epsilon_{\rm C}=0.15 \epsilon_{\rm C,\odot}$, 
$\epsilon_{\rm O}=0.3 \epsilon_{\rm O,\odot}$ are obtained from
O\,{\sc ii} $\lambda\lambda$4097-4120, 
C\,{\sc iii} $\lambda\lambda$4647-51 and 
O\,{\sc  ii} $\lambda\lambda$4638-76 (a blend),
plus other O\,{\sc ii} features.

In the UV, comparison between 
observations and synthetic spectra is hindered by the low S/N in the 
{\iue} HIRES dataset. Nevertheless, the 
very strong Fe\,{\sc iv} absorption between
$\lambda\lambda$1500 and 1750 
is well matched by the synthetic spectrum 
(see Fig.~\ref{sk66169_cmfgen3}). Si\,{\sc iv}
$\lambda\lambda$1393--1402 is well matched, as is 
Al\,{\sc iii} $\lambda\lambda$1852-62, although 
blanketing  
produces a C\,{\sc iv} $\lambda\lambda$1548--51 P Cygni  emission profile which
is clearly too weak in the synthetic spectrum. The only other prominent
feature not matched by the model is (presumably) N\,{\sc v} 
$\lambda\lambda$1238-42, which is not predicted with our assumed
X-ray parameters. The only discernable effect of X-rays is a slight 
strengthening of C\,{\sc iv} $\lambda\lambda$1548--51.

For the {\fuse} FUV dataset, 
agreement is excellent for $T_{\rm eff}$=26kK, namely
P\,{\sc iv} $\lambda$950,
P\,{\sc iv} $\lambda$950 and $\lambda\lambda$1025-1033, 
N\,{\sc ii} $\lambda\lambda$1084-6, S\,{\sc iv} 
$\lambda\lambda$1062--1073, 1099 and  Si\,{\sc iii} $\lambda\lambda$1110-1113. 
N\,{\sc ii} is one of the few 
abundance-sensitive UV lines, although
it is strongly temperature sensitive, which
hinders its reliable use. Our assumed X-ray model has no effect on
FUV wind diagnostics.

Amongst other UV and FUV lines,  
only Al\,{\sc iii} $\lambda$1852 (fit worsens) and C\,{\sc iii}
$\lambda$977 (fit improves) are particularly sensitive to a reduced
temperature of $T_{\rm eff}$=24kK, as implied by the
He\,{\sc ii} $\lambda$4686 emission strength.


\section{Abundances}\label{abundances}

Our results are amongst the
first determinations of CNO abundances in O supergiants, although
AB supergiants have been studied in detail (Venn 1996, 1999).

Metal abundance studies of Galactic O supergiants remain 
surprisingly sparse: 
Pauldrach et al. (1994, 2001) obtained partially CNO-processed material for 
$\zeta$ Pup (O4\,I(n)f) and $\alpha$ Cam (O9.5\,Ia)
with N/C$\sim$5 by number, 
whilst Villamariz et al. (2002) recently obtained N/C=1 for HD\,209975 (O9.5\,Ib).
Taresch et al. (1997)
studied the nitrogen spectrum of HD\,93129A (O2\,If$^{\ast}$) revealing
$\epsilon_{\rm N}=2 \epsilon_{\rm N,\odot}$, whilst
Pauldrach et al. (2001) determined N/C$\sim$4 by number for
 $\alpha$ Cam 
(O9.5\,Ia). 

Amongst Magellanic Cloud O stars, the sole result involving
carbon and nitrogen was by Haser et al. (1998) who obtained N/C$\sim$5
for the SMC O2 giant NGC\,346 \#3 (Walborn et al. 2002b). Haser et al.
(1998) were unable to determine CNO abundances for {\ofour} due to problems
with reproducing the unsaturated lines of N\,{\sc iv} $\lambda$1718,
O\,{\sc iv} $\lambda\lambda$1338-43 and O\,{\sc v} $\lambda$1371. Such
problems were probably due to the use of an erroneous stellar temperature
for {\ofour}, and the fact that (more sensitive) optical lines were
excluded.

Our results broadly support previous investigations, namely (substantial)
nitrogen enrichment, plus (modest) carbon depletion, i.e.
reveal N/C$\sim$3 (\onine), 10 
(\osix, \oseven) or 30 (\ofour), by number. Unfortunately, Venn (1999) was
only able to derive an upper limit to the carbon abundance in her study 
of SMC A supergiants, such that N/C$\ge$1. Nevertheless, she also supported
a general nitrogen enrichment, of between 0.2 and $\ge$1.2 dex. 
How do such abundance ratios compare with  theoretical expectations for
post-main sequence massive stars?

A major deficiency with the previous generation of evolutionary models
for massive stars 
was that He and N enrichments 
for OB stars were not predicted until much later stages of 
evolution (Maeder \& Meynet 2000). 
In contrast, recent models  accounting for rotational 
mixing do predict N/C ratios which are much higher than the initial
(solar)
N/C ratio of $\sim$0.2 by number. For example, 
Meynet \& Maeder (2000) discuss the evolution of an initial 
60 $M_{\odot}$ star at 
(solar)
metallicity, initially rotating at 300~km\,s$^{-1}$. Rotational mixing permits
changes to the surface N/C ratio, i.e. N/C=6 during the O-supergiant
phase, when He/H=0.2, and subsequently N/C=12 when He/H=0.25 (Meynet,
priv. comm.).

In general, our results indicate moderate carbon and oxygen
depletions  relative to LMC/SMC H\,{\sc ii} regions, with nitrogen
substantially enriched. If we were to assume that we are solely witnessing
CNO-processed material at the surface, 
one would expect carbon (and oxygen) depletions which are 
substantially greater to produce the necessary nitrogen enrichment,
since the total CNO abundance is maintained within the CNO-cycle. 
Indeed, the problem is potentially even more acute, since 
up until now we have adopted nominal
LMC (SMC) CNO abundances which are scaled to 0.4 (0.2) $\epsilon_{\odot}$,
where $\epsilon_{\odot}$ refers to the solar metallicity of a particular
element.
In reality, the Magellanic Cloud 
nitrogen abundance from which these stars recently formed
is considered to be much more depleted,
with $\sim$0.1$\epsilon_{\odot}$ in the LMC (Garnett 1999; Korn
et al. 2002), and 0.03$\epsilon_{\odot}$ in the SMC (Peimbert, Peimbert
\& Ruiz 2000). 
  
The apparent problem is best illustrated for \oseven\ in the SMC for which 
carbon is only a factor of two depleted  relative 
to  typical SMC values (Garnett et al. 1995), whilst 
nitrogen is a factor of one hundred 
times higher than  normal SMC values. 
Oxygen cannot easily remedy this situation, since the 
CN cycle precedes the ON cycle. 
Besides, it suffers a similar depletion to carbon (Peimbert et al. 2000). 
Similarly, taking \osix\ as representative of the 
LMC stars, discrepancies are 
somewhat similar: carbon and oxygen are again only a factor of two 
depleted relative to H\,{\sc ii} regions
or B stars, whilst nitrogen is a factor of 50 times
higher than LMC values (Garnett 1999; Korn et al. 2002).

How can we resolve this puzzle?
A likely solution is that we are witnessing evidence
of mixing. The surface abundances are not solely exhibiting products
of CNO processing, but instead a mixture of initial abundances, plus
CNO processed material. In this case we would expect strong N enrichment,
which is observed, but (assuming, say, 50\% initial abundances) a 
reduction in C by no greater than a factor of two, which we also observe.
In principal, one could get an estimate of the amount of mixing that has 
occured.
Detailed comparisons with evolutionary models
accounting for rotation await appropriate calculations for
low-metallicity environments. 
One might question the high nitrogen abundance in 
\oseven~ as determined by {\sc cmfgen}, but 
N\,{\sc iii} $\lambda$4097 appears to be a very sensitive abundance indicator. 
Indeed, the optical N\,{\sc iii} spectral morphology of 
\oseven\
is remarkably similar to Galactic counterparts such as HD\,163758 (O6.5\,Iaf),
suggesting similar 
present-day N abundances,
despite very different initial abundances.

\section{Clumping in O supergiants?}\label{clumping}

Observations universally point to the structured nature of OB winds, via
UV (e.g. Massa et al. 1995) and optical 
spectroscopic monitoring campaigns (e.g. Eversberg,
Lepine \& Moffat 1998), and mm/radio imaging (e.g. Blomme et al. 2002).
Hydrodynamic
models also 
indicate that their winds are clumped and time variable (e.g. Owocki
et al. 1988).
For Wolf-Rayet stars, all stellar lines 
are formed in the clumped 
wind whilst the situation for O-type stars is less clear, with clumping 
originating above the stellar surface -- perhaps above the 
formation region of the usual diagnostic lines.  
It is apparent that shocked, clumpy, models {\it are} 
required to match the O\,{\sc vi} doublet in our program stars, plus
N\,{\sc v} in some cases. 
Precise shock  parameters 
cannot yet be derived uniquely from optical and UV spectra. 
Nevertheless,  it is possible 
that clumping may also affect some of our adopted wind diagnostics of O stars.
If so, we may use this fact to investigate the degree of clumping for the
inner wind of O stars. We have calculated a model of {\oseven} identical to 
that discussed above, except that its volume filling factor is reduced from
100\% to 10\%, and mass-loss rate reduced to 
2$\times 10^{-6}$ $M_{\odot}$ yr$^{-1}$. Overall, we find very few 
differences from the smooth model. The O supergiant 
case is in stark contrast with Wolf-Rayet
stars for which electron scattering wings provide very important probes of
clumping (Hillier 1991; Schmutz 1997).

One potentially useful diagnostic is the P\,{\sc v} $\lambda\lambda$1118-28
doublet for which the
non-clumped model 
overpredicts both
absorption and emission for {\ofour}, {\osix} and {\oseven}
(see also Massa et al. 2002).  
P$^{4+}$ 
is the dominant phosphorus ion throughout most of
the wind of {\oseven} (if $T_{\rm eff} \sim$ 32kK), so that 
the clumped model reveals weaker profiles, 
in better agreement with observation - see Fig.~\ref{av232_pv}. 
One {\it might} argue that P\,{\sc v} represents a useful
probe of the degree of clumping 
in O stars, {\it except} that a 
similar effect is found by reducing the elemental abundance of 
phosphorus from $\epsilon_{\rm P}=0.2 \epsilon_{\rm P,{\odot}}$ to
$\epsilon_{\rm P}=0.05 \epsilon_{\rm P,{\odot}}$
(Fig.~\ref{av232_pv}). Similar results are obtained for {\ofour} and {\osix}.
Pauldrach
et al. (1994) have previously indicated a reduced abundance of phosphorus
by a factor of 1.5--2 relative to 
solar values
for $\zeta$ Pup (O4\,I(n)f) 
from P\,{\sc v} $\lambda\lambda$1118--28 observations with {\it Copernicus}.

\section{Discussion and Conclusions}\label{conclusions}

Our analysis of optical H-He wind and photospheric profiles of luminous
O supergiants, using line-blanketed, extended model atmospheres, 
reveal systematically ($\sim$15--20\%) lower temperatures
than plane-parallel results based solely on optical H-He photospheric lines
(e.g. Herrero et al. 1992). Our results are supported by 
UV {\hst} and especially FUV {\fuse} spectroscopy of metal wind lines and
 photospheric iron lines in these stars. 
Initial {\fuse} datasets of {\oseven} 
raised questions about
the validity of
temperatures derived from plane-parallel O supergiant models (Paper~I),
which unified model atmospheres such as Hillier \& Miller (1998) 
can now resolve.

Fig.~\ref{osuper} compares temperatures
for O supergiants derived here with those from previous compilations by 
B\"{o}hm-Vitense (1981),
Schmidt-Kaler (1982), Howarth \& Prinja (1989) and most recently
by Vacca et al. (1996).
Of these, one might expect Vacca et al. to be
the closest match to the present results, yet the reverse is true, especially
at the earliest subtypes. This can be understood from the fact that Vacca
et al. included the most recent spectroscopic results, {\it excluding} wind
or line blanketing. Consequently, $\zeta$ Pup (O4\,I(n)f) was 
 included in their calibration with $T_{\rm eff}$(no wind blanketing)=46.5kK, 
rather than $T_{\rm eff}$(wind blanketing)=42kK (Bohannan et al. 1986),
although the `core-halo' approach was followed in their study. Test
calculations carried out  for $\zeta$ Pup, which consistently include
sphericity {\it and} line blanketing, indicate $T_{\rm eff}\sim39$kK,
illustrating the additional effect that line blanketing plays.

More recently, Herrero et al. (2000) compared results of early O supergiants
obtained with plane-parallel models that accounted for line blanketing
with those obtained using spherical models for which blanketing was omitted.
For HD\,15570 (O4\,If$^+$), Herrero et al. obtained $T_{\rm eff}$(blanketed)=50kK
versus $T_{\rm eff}$(spherical)=42kK. Stellar temperature determinations
of O supergiants undertaken during the past decade are summarised in 
Table~\ref{t6}, sorted by the degree of complexity implemented; i.e.,
plane-parallel versus spherical, plus
unblanketed versus line blanketed. From Table~\ref{t6}, blanketing {\it 
and} sphericity have rarely
previously been simultaneously considered (Pauldrach et al. 1994, 2001),
with only Taresch et al. (1997) successfully combining optical/UV/FUV
diagnostics.  A critical revision to the temperature calibration of O 
supergiants clearly requires analysis of a substantially larger sample of
targets, using the methods outlined here, which is currently underway. 
Revisions do not necessarily possess a strong metallicity dependence, 
since {\oseven} (SMC) possesses a similar difference from standard
calibrations to the three LMC stars studied
here (Fig.~\ref{osuper}). Wind strength is more critical -- one would
expect the greatest deviations from conventional temperature scales
for those stars with the highest wind densities, i.e. those with 
H$\alpha$ and He\,{\sc ii} $\lambda$4686 emission.

Similar calculations for O dwarfs -- also allowing for line blanketing -- 
obtained $\sim$5\% lower temperatures than unblanketed results
(Herrero, Puls \& Villamariz 2000; Martins et al. 2002). 
Independent observational evidence in 
favour of  (1--2kK) lower temperatures of O dwarfs may be drawn 
from studies of O-type  binaries (e.g. Harries, Hilditch \& Hill 1998).
The rarity of O supergiants within short-period eclipsing binary systems
prevents direct determinations of radii, and thus temperatures, for 
O supergiants.
The greater effect identified here for extreme O supergiants may reasonably
be attributable to the effect of strong winds on the photospheric lines.

Our results argue for a substantial revision in stellar parameters 
for O supergiants 
versus those derived from previous spectroscopic techniques.
For {\ofour}, the reduction in temperature implies a decrease in luminosity 
from $\sim$1.6$\times 10^{6} L_{\odot}$ to 1.0$\times 10^{6} L_{\odot}$.
Such changes greatly affect the number of 
Lyman-ionizing photons 
emitted
-- in the case of {\ofour}, the standard 
$T_{\rm eff}$-calibration would imply that its 
Lyman-ionizing output
is a  factor of two higher than 
the
10$^{49.7}$ ph\,s$^{-1}$ determined here. 
The ionizing flux of {\ofour} below the He\,{\sc i} $\lambda$504 edge is 
reduced by a factor 
of three to 10$^{49.0}$ ph\,s$^{-1}$.  Similar changes
are obtained for the other program stars. 
Fortunately, comparisons of nebular strengths in young, massive clusters 
with their constituent O stars are generally weighted towards 
main-sequence populations. 
Nevertheless, cases exist where individual O
stars dominate H\,{\sc ii} regions (e.g. Oey et al. 2000) which would be
dramatically affected by such large temperature changes.

Masses are more difficult to constrain, but the decrease in 
the spectroscopic luminosity and gravity also cause large differences 
from
previous determinations. Consequently, evolutionary model
comparisons with O supergiants carried out previously may not have been
using the appropriate tracks in many cases. Although initial 
(and indeed current) rotational velocities for the program stars are not 
known, adopting $v_{\rm init}$=300 km\,s$^{-1}$ suggest an 
{\it initial}  mass
of 75$M_{\odot}$ for {\ofour} according to evolutionary models (Fig.~8 of 
Meynet \&  Maeder 2000). In contrast,
an initial mass in excess of 120$M_{\odot}$ would
be implied by models on the basis of its previously
determined higher luminosity. Our results do not solve the 
long-standing `mass discrepancy' for O stars (Herrero et al. 1992) between
evolutionary and spectroscopic mass determinations,  since 
the reduction in spectroscopic luminosity is also 
accompanied by a reduced mass. Again, using {\ofour} as an example,
the reduction in spectroscopic gravity from $\log g\sim$3.7 to $\log 
g\sim$3.35 implies a decrease from $\sim70M_{\odot}$ to 
$\sim35M_{\odot}$ in its {\it current} mass.

The revision in luminosity also 
implies a different mass-loss rate. Table~\ref{t4} shows a 
35\% reduction in $\dot{M}$ for {\ofour}
(without accounting  for the  possibility that the wind is clumped).
This revision affects the kinetic energy 
injected by individual stars or a young, massive cluster to the ISM,
since this is more dependent on the O 
supergiants and WR properties than on the main sequence stars 
(e.g. Crowther \& Dessart 1998).

These effects have 
major consequences for the Wind-Momentum-Luminosity Relationship (WLR).
In Fig.~\ref{momentum}, we present the WLR for 
Galactic (circles), 
LMC (squares) and SMC (triangles)
luminous O stars from
Puls et al. (1996, open symbols). Also shown is the form of the 
Wind-Momentum-Luminosity Relationship
(dotted line) for Galactic O supergiants according to Kudritzki \& Puls (2000).
We have added current results for our sample of LMC and SMC targets
(filled symbols),
 two of which are in common with Puls et al. 
(1996). Revisions in parameters are illustrated in these cases with
arrows. Clearly, when the present results are 
extended to
a larger sample of luminous OB stars, substantial revisions to the empirical
relationship of Kudritzki \& Puls (2000) are expected.


FUV {\fuse} spectroscopy has proved to be  invaluable for the 
present analysis. In contrast with {\hst}/{\iue}, the availability of
unsaturated resonance lines from dominant ionization stages 
of ``cosmically rare'' elements (e.g.
P\,{\sc v} $\lambda\lambda$1118--28) represents a potentially 
exciting new diagnostic of wind clumping in  O stars. Elements such as
S and P are especially useful since they do not change substantially
dueing the evolution of a massive star, in sharp contrast to CNO. 
{\fuse} provides
many additional wind features, such that we no longer have access to
only the saturated (model insensitive) C\,{\sc iv} $\lambda\lambda$1548--51
and X-ray influenced N\,{\sc v} $\lambda\lambda$1238--42 wind lines.  
Our ongoing {\fuse} program will aim to undertake 
detailed modeling of UV wind lines for a large sample of O stars, 
covering a greater range  of spectral types and luminosity classes.


\acknowledgments
This work is based, in part, 
on data obtained by
the NASA-CNES-CSA {\fuse} mission operated by the Johns Hopkins University.  
Financial support has been provided by NASA contract NAS5-32985 
(U. S. participants), the Royal Society (PAC) and PPARC (OD, CJE). 
We wish to thank John Hutchings and Ken Sembach for the use of data
from P.I. Time programs P117 and P103, respectively;
Adi Pauldrach, Alex de Koter for the use of their
stellar-atmosphere codes; 
and Goetz Gr\"{a}fener for undertaking test
calculations for $\zeta$ Puppis. We are grateful to Nolan Walborn
for detailed comments  on an draft of this paper.

\begin{small}
\begin{table*}
\caption{Target Stars. Assumed distance moduli to the Magellanic Clouds
are 18.5 (LMC) and 18.9 (SMC), whilst reddenings 
are derived from the mean of (i)
intrinsic colours from FitzGerald (1970) and Schmidt-Kaler (1982); (ii) fitting
final stellar models to UV-optical spectrophotometry. Atomic and molecular 
hydrogen column
densities are obtained from fitting Ly$\alpha$ in this work, and fitting
H$_{2}$ lines in {\fuse} spectra by Tumlinson et al. (2002), respectively.}
\label{t1}
\begin{center}
\begin{small}
\begin{tabular}{
l@{\hspace{2mm}}
l@{\hspace{2mm}}
c@{\hspace{2mm}}
c@{\hspace{2mm}}
c@{\hspace{2mm}}
c@{\hspace{2mm}}
c@{\hspace{2mm}}
c@{\hspace{2mm}}
c@{\hspace{2mm}}
c@{\hspace{2mm}}
c
@{\hspace{2mm}}
c}
\tableline\noalign{\smallskip}
Object                     &
Alias                         &
Galaxy                     &
Sp. Type              &
Ref.                       &
$V$                        &
$B-V$                      &
Ref.                       &
$E(B-V)$                  &         
$\log N$(H\,{\sc i})             &
$\log N$(H$_{2}$)               &
$M_{\rm V}$               \\
 & & & & & mag & mag & & mag & cm$^{-2}$ & cm$^{-2}$ & mag \\
\tableline\noalign{\smallskip}
   \ofour & Sk-67$^{\circ}$ 166 & LMC & O4\,Iaf$+$ & 1 & 12.27 &$-$0.22 & 2 & 0.08 &20.7&   15.7& $-$6.5\\
   \osix & Sk-65$^{\circ}$ 22 & LMC & O6\,Iaf$+$ & 1 & 12.07 &$-$0.19 & 3 
& 0.07   &21.0&   14.9& $-$6.7\\
   \oseven & Sk 80 & SMC & O7\,Iaf$+$ & 1 & 12.36 &$-$0.21 & 5 & 0.09    
&21.0&   15.3& $-$6.8\\
   \onine &        & LMC & O9.7\,Ia$+$& 4 & 11.56 &$-$0.13 & 4 & 0.10    
&20.5&$<$14.0& $-$7.3\\
\tableline\noalign{\smallskip}
\end{tabular}
  (1) Walborn 1977;
  (2) Ardeberg et al. 1972; 
  (3) Isserstedt 1979;
  (4) Fitzpatrick 1988;
  (5) Azzopardi, Vigneau \& Macquet 1975
\end{small}
\end{center}
\end{table*}
\end{small}

\clearpage

\begin{footnotesize}
\begin{table*}
\caption[]{
Wind velocities ($v_{\rm black}$) observed at UV ({\iue}, {\hst}) and 
FUV ({\fuse}) wavelengths for our program stars, including
radial velocity measurements, $v_{r}$.  Previous determinations are
by Haser (1995, H95) and by Massa et al. (2002, M02). Our adopted
value is provided in the final column, taken from N\,{\sc iii}, with
the exception of {\onine} for which Si\,{\sc iv} was used. 
}
\label{t2}
\begin{center}
\begin{small}
\begin{tabular}{
l@{\hspace{2mm}}
l@{\hspace{2mm}}
l@{\hspace{2mm}}
r@{\hspace{2mm}}
r@{\hspace{2mm}}
r@{\hspace{2mm}}
r@{\hspace{2mm}}
r@{\hspace{2mm}}
r@{\hspace{2mm}}
r@{\hspace{2mm}}
r}
\tableline\noalign{\smallskip}
Object&Sp. Type&Dataset&$v_{r}$&
\ion{N}{3} &
\ion{N}{5} & 
\ion{Si}{4}&
\ion{C}{4} &
H95 & M02 &
{\vinf} \\
 & & & km\,s$^{-1}$ &
$\lambda$989.80  & 
$\lambda$1238.82 & 
$\lambda$1393.76 & 
$\lambda$1548.19 & 
km\,s$^{-1}$ &
km\,s$^{-1}$ &
km\,s$^{-1}$ 
\\
\tableline\noalign{\smallskip}
   \ofour & O4~Iaf$+$ &{\fuse,\hst} &$-$265&1750  &1740 &     &1725 & 1900
& 1800 & 1750 
\\
   \osix & O6~Iaf$+$ &{\fuse,\iue}  &$-$240& 1520 &     &1460 &1520 &     
& 1350 & 1520 
\\
   \oseven & O7~Iaf$+$ &{\fuse,\hst}&$-$165& 1330 &1360 &1290 &1225 & 1400
&    & 1330 
\\
   \onine & O9.7~Ia$+$&{\fuse,\iue} &$-$295&  910 &     &1000 & 990 & 
& 800 & 1000\\
\noalign{\smallskip}
\tableline
\end{tabular}
\end{small}
\end{center}
\end{table*}
\end{footnotesize}

\clearpage

\begin{small}
\begin{table*}
\caption[]{Stellar parameters of program stars derived here
from hydrostatic, plane-parallel models ({\sc tlusty}), together with previous
determinations from Puls et al. (1996) or Lennon et al. (1997).
Radii (and hence luminosities) are derived following $5\log 
(R/R_{\odot})=29.57-M_{\rm V} - {\rm V}(T_{\rm eff})$ from 
Herrero et al. (1992) for consistency.}
\label{t3}
\begin{center}
\begin{tabular}{llcccccccc}
\tableline\noalign{\smallskip}
Object& Sp Type & $T_{\rm eff}$(kK) & $R/R_{\odot}$ & 
$\log(L/L_{\odot}$) & $\log g$ & $y$ 
& $M(M_{\odot}$) & $v \sin i$ & Ref.\\
\tableline\noalign{\smallskip}
   \ofour  & O4~Iaf$+$  & 46.5  & 20.6  &  6.25&   3.7  &0.09& 70 
&100&This 
work\\
           &            & 47.5  & 19.5 &  6.24&   3.6  &0.1 &    & 
80&Puls et al.\\  
   \osix   & O6~Iaf$+$  & 41    & 25.6   &  6.22&     3.5 &0.09& 68 
&90&This 
work   \\
   \oseven & O7~Iaf$+$  & 39   & 26.3  &  6.16&     3.4 &0.12& 57& 60&This 
work\\
           &            & 37.5 & 29.5  &  6.19&     3.2 &0.2&  &  &Puls et 
al.\\
   \onine  & O9.7~Ia$+$ & 31   &38.5   &  6.09&     3.1 &0.09& 27&100&This 
work\\
           &            & 28   & 44.6  &  5.98&         &    &&&Lennon et 
al.\\
\noalign{\smallskip}
\tableline
\end{tabular}
\end{center}
\end{table*}
\end{small}

\clearpage


\begin{small}
\begin{table*}
\caption[]{Atomic models for early and late O supergiants. For each ion 
F denotes full levels, S super levels, and T the 
number of bound-bound transitions considered. Most ions are common
to all stars, except where noted: C\,{\sc ii}, O\,{\sc ii}, Al\,{\sc iii}, Si\,{\sc iii}, S\,{\sc iii} 
and Fe\,{\sc iii} are only included in late O supergiant models (indicated in bold); 
whilst N\,{\sc v}, O\,{\sc vi}, S\,{\sc vi} and Fe\,{\sc vii} are only included in early O supergiant
models (shown in italics).}
\label{table0}
\begin{center}
\begin{tabular}{l@{\hspace{-5mm}}
c@{\hspace{1mm}}c@{\hspace{1mm}}c
c@{\hspace{1mm}}c@{\hspace{1mm}}c
c@{\hspace{1mm}}c@{\hspace{1mm}}c
c@{\hspace{1mm}}c@{\hspace{1mm}}c
c@{\hspace{1mm}}c@{\hspace{1mm}}c
c@{\hspace{1mm}}c@{\hspace{1mm}}c
c@{\hspace{1mm}}c@{\hspace{1mm}}c}
\tableline
Element & 
\multicolumn{3}{c}{I} & 
\multicolumn{3}{c}{II} & 
\multicolumn{3}{c}{III} & 
\multicolumn{3}{c}{IV} & 
\multicolumn{3}{c}{V} & 
\multicolumn{3}{c}{VI} & 
\multicolumn{3}{c}{VII}\\
        &F & S & T 
	&F & S & T 
	&F & S & T 
	&F & S & T
	&F & S & T
	&F & S & T
	&F & S & T \\
\tableline
H       & 30 & 20 & 435 \\
He      & 49 & 40 & 325 & 30 & 20 & 435 \\
C       &    &    &     & {\bf 30}&{\bf 16} & {\bf 115}& 54& 29 & 268 
                        & 18      & 13      & 76       \\
N       &    &    &     & {\bf 41}&{\bf 21} & {\bf 144}& 90 & 47& 733   
                        & 60      & 34      & 331      
                        & {\it 18}&{\it 13} &{\it 85}  \\
O       &    &    &     & {\bf 60}&{\bf 20} & {\bf 405}& 45 & 25& 182 
                        & 29      &     13  & 148      & 23 & 13&  65 
                        &{\it 15} &{\it 9}  &{\it 42}  \\
Al      &    &    &     &    &    &     &{\bf 65} & {\bf 21}& {\bf 1452} \\
Si      &    &    &     &    &    &     &{\bf 45} & {\bf 25}& {\bf 172} &  33 & 23 & 185 \\
P       &    &    &     &    &    &     &    &   &     & 28 & 16 & 57 & 28& 18 & 139 \\
S       &    &    &     &    &    &     &{\bf 41}  & {\bf 21}&  {\bf 177}&      92 &    37 & 708 
                                        &44        & 24      & 163       & {\it 11}&{\it 7}&{\it 24}\\ 
Fe      &    &    &     &    &    &     &{\bf 607} & {\bf 65}& {\bf 5482}&      272&    48 &3113 
                                        & 182      & 46      &1781       &      270 &   36 &2858
                                        & {\it 153}& {\it 29}&{\it 1095} \\
\tableline
\end{tabular}
\end{center}
\end{table*}
\end{small}

\clearpage

\begin{table*}
\caption[]{Stellar models for {\ofour} (O4\,Iaf$^+$), selected
to match the observed H$\alpha$ emission and absolute magnitude, $M_{\rm V}$
($h$=0.005 $R_{\odot}$). In 
all cases, two models are calculated, one with CNO elemental abundances
set to $\epsilon=0.4\epsilon_{\odot}$, with helium and nitrogen 
assumed to be enriched in the other (He/H=0.2 by number,
$\epsilon_{\rm N}=6 \epsilon_{\rm N,\odot}$) with carbon and 
oxygen both depleted ($\epsilon_{\rm C}=0.05\epsilon_{\rm C,\odot}$,
$\epsilon_{\rm O}=0.1\epsilon_{\rm O,\odot}$).
A supersonic velocity law with $\beta$=1.0 is found to provide the best
match to the shape of H$\alpha$.}
\label{t4}
\begin{center}
\begin{tabular}{llcccccccc}
\tableline\noalign{\smallskip}
$T_{\rm eff}$ &$R_{\ast}$ &$\log(L/L_{\odot}$)&$\log g$&$\dot{M}$ &$v_{\infty}$\\
   kK         &$R_{\odot}$&                   &cgs&$M_{\odot} {\rm yr}^{-1}$ 
& km\,s$^{-1}$\\
\tableline\noalign{\smallskip} 
 46  &  19.1 &  6.17 & 3.55 & 9.75$\times 10^{-6}$ & 1750 \\ 
 43  &  19.7  & 6.08 & 3.45 & 9.75$\times 10^{-6}$ & 1750 \\ 
 40  & 20.5 &  5.98  & 3.35 & 8.5$\times 10^{-6}$ & 1750 \\ 
 37 &  21.0 & 5.88 & 3.25 & 6.75$\times 10^{-6}$ & 1750 \\ 
 34 & 22.3 & 5.78 & 3.15  & 7.25$\times 10^{-6}$ & 1750 \\ 
\noalign{\smallskip}
\tableline
\end{tabular}
\end{center}
\end{table*}

\begin{small}
\begin{table*}
\caption[]{Stellar parameters of program stars derived from stellar wind models
(Hillier \& Miller 1998), including bolometric corrections, B.C. 
Approximate elemental mass fractions, $\epsilon$,
relative to solar  values are indicated, 
whilst He/H=0.2 is adopted for all program stars.
Previous wind determinations (based in part on 
hydrostatic models) are  also 
included here for comparison (Puls et al. 1996; Lennon et al. 1997).
}
\label{t5}
\begin{center}
\begin{small}
\begin{tabular}{l@{\hspace{2mm}}
l@{\hspace{2mm}}c@{\hspace{2mm}}
c@{\hspace{2mm}}c@{\hspace{2mm}}
c@{\hspace{2mm}}
c@{\hspace{2mm}}r@{\hspace{2mm}}
c@{\hspace{2mm}}r@{\hspace{2mm}}
c@{\hspace{2mm}}c@{\hspace{2mm}}c@{\hspace{2mm}}l}
\tableline\noalign{\smallskip}
Object& Sp Type   & $T_{\rm eff}$     &  $R_{\ast}$   &
 $\log g$   &B.C.&  $\log(L/L_{\odot}$)   &  $\dot{M}$ & $\beta$ & {\vinf}  
& 
$\epsilon_{C}$ & $\epsilon_{N}$ & $\epsilon_{O}$ &
Ref.    \\
     &           &   kK   & $R_{\odot}$ & cgs & mag & & $M_{\odot} {\rm 
yr}^{-1}$ && 
km s$^{-1}$ & $\epsilon_{C,\odot}$ & $\epsilon_{N,\odot}$ &  $\epsilon_{O,\odot}$ & \\
\tableline\noalign{\smallskip}
   \ofour  & O4~Iaf$+$  & 40 & 20.5 & 3.6 & $-$3.7 & 5.98 &8.5$\times 
10^{-6}$ & 
1.0& 1750&0.05&9.0&0.1&This work\\ 
           &            & 47.5&19.5 & 3.6 && 6.24&13.0$\times 10^{-6}$& 
0.7& 1900 &    &   &  &Puls et al.\\
   \osix   & O6~Iaf$+$  & 33.5 & 25.0 & 3.2 &$-$3.2& 5.86 &11.0$\times 
10^{-6}$ & 
1.3 & 1520&0.1 &4.5&0.2&This work\\
    \oseven & O7~Iaf$+$ & 32 & 27.5 & 3.1 &$-$3.1 & 5.85 & 4.5$\times 
10^{-6}$ & 
1.65 & 1330 & 0.07 &2.0 & 0.1 & This work \\ 
           &            & 37.5&29  & 3.2 & &6.19&5.5$\times 10^{-6}$ &1.4 
& 1400  &    &   &  &Puls et al.\\
     \onine & O9.7~Ia$+$ &
 26 & 40  & 2.8 & $-$2.5 & 5.82 & 6.0$\times 10^{-6}$ & 1.75 & 1000 & 0.15 
& 2.0 & 0.3 & This work \\ 
           &            & 28 & 41.5&     && 6.00 &20.0$\times 10^{-6}$ &-- 
& 1000  &    &   &  &Lennon et al.\\
\noalign{\smallskip}
\tableline
\end{tabular}
\end{small}
\end{center}
\end{table*}
\end{small}

\clearpage
\begin{table*}
\caption[]{Stellar temperatures (in kK) of 
O supergiants as derived from plane-parallel (p-p) models including/excluding
line blanketing (l.b.) and/or sphericity (sph). 
Wind blanketed results from the core-halo approach are presented in
parenthesis, whilst spherical, l.b. models using optical, UV 
{\it and} FUV diagnostics are indicated in bold.}
\label{t6}
\begin{center}
\begin{small}
\begin{tabular}{lll@{\hspace{5mm}}l@{\hspace{8mm}}ll@{\hspace{8mm}}lc}
\tableline\noalign{\smallskip}
Object& Sp Type & galaxy &
 $T_{\rm eff}^{\rm p-p}$
& $T_{\rm eff}^{\rm p-p+l.b.}$
& $T_{\rm eff}^{\rm sph}$  
& $T_{\rm eff}^{\rm sph+l.b.}$& Ref \\ 
\tableline\noalign{\smallskip}
HD\,93129A & O2~If*   & Galaxy     & 50.5 &    &   & {\bf 52} &5,6   \\
Cyg OB2 \#7& O3~If*   & Galaxy     &      & 51 &   &    &8   \\ 
   \ofour  & O4~Iaf$^+$& LMC       & 49 &    &   & {\bf 40} &11 \\
Sk\,$-67{\arcdeg}167$ & O4~Inf$^+$& LMC & 47.5 &    &   &    &5 \\
HD\,15570  & O4~If$^+$ & Galaxy    &      & 50 & 42&    & 8  \\
$\zeta$ Pup& O4~If     & Galaxy    & 46.5 &    &(42)& 42 & 1, 4  
\\
HD\,14947  & O5~If$^+$ & Galaxy    & 43.5 & 45 & 40&    & 5,8   \\
Cyg OB2 \#8c& O5~If    & Galaxy    &      & 48 &   &    & 7   \\
Cyg OB2 \#9& O5~If     & Galaxy    &      &44.5&   &    & 7 \\
Cyg OB2 \#11& O5~If$^+$& Galaxy    &      & 43 &   &    & 7    \\ 
   \osix   & O6~Iaf$^+$& LMC       & 41   &    &   &{\bf 33.5}& 11 \\
$\lambda$ Cep & O6~I(n)fp& Galaxy  & 38   &41.5& 37&    & 5,8  \\
   \oseven & O7~Iaf$^+$ & SMC      & 39 &    &   & {\bf 32} & 11 \\
HD~192639  & O7~Ib(f)  & Galaxy    & 38.5 &    &   &    & 3 \\
HD~193514  & O7~Ib(f)  & Galaxy    & 38   &    &   &    & 3     \\
HD~210809  & O9~Ib     & Galaxy    & 33   &    &   &    & 3 \\
AzV 469     & O9~Ib    & SMC       & 34   &    &   &    & 9 \\ 
$\alpha$ Cam& O9.5~Ia  & Galaxy    & 30   &    &(30)& 29   & 2,5,10\\ 
Cyg OB2 \#10& O9.5~I   & Galaxy    &      &31  &   &    & 7   \\
HD~209975   & O9.5~Ib  & Galaxy    &32.5  &    &   &    & 3 \\
HD~19409    & O9.7~Ib  & Galaxy    & 31.5 &    &   &    & 3 \\
   \onine  & O9.7~Ia$^+$ & LMC     & 31   &    &   & {\bf 26} & 11 \\
\noalign{\smallskip}
\tableline
\end{tabular}
\end{small} 
\end{center}
\begin{footnotesize}
(1) Bohannan et al. (1986); (2) Voels et al. (1989); 
(3) Herrero et al. (1992); 
(4) Pauldrach et al. (1994);
(5) Puls et al. (1996); (6) Taresch et al. (1997);
(7) Herrero et al. (1999); (8) Herrero et al. (2000); 
(9) Dufton et al. (2000); (10) Pauldrach et al. (2001); (11) this work
\end{footnotesize}
\end{table*}


\clearpage

\begin{figure}[h!]
    \epsscale{1.0}
    \caption{
     Calibrated {\fuse} spectra of {\ofour}, {\osix}, {\oseven}, {\onine} (top
     to bottom). Important stellar features are identified in the 
     panels. Lines due to the atomic Lyman series are indicated.
     \label{fusedata}
     } 
\end{figure}

\begin{figure}[h!]
    \epsscale{1.0}
    \caption{
     Ultraviolet spectrograms of our program O supergiants based on 
     {\iue} or {\hst}
     observations.
     \label{iue-hst}
     } 
\end{figure}

\begin{figure}[h!]
    \epsscale{1.0}
    \caption{
     UVES H$\alpha$ profiles for our program O supergiants.
     \label{ha-atlas}
     } 
\end{figure}


\begin{figure}[h!]
    \epsscale{1.0}
    \caption{
     Comparison between UVES observations (solid) of 
     {\ofour} (O4\,Iaf$^+$) and synthetic profiles      (red, dotted) 
     obtained with {\sc tlusty} ($T_{\rm eff}$=46.5kK, $\log g$=3.7).
     \label{hde269698_tlusty}
     } 
\end{figure}

\begin{figure}[h!]
    \epsscale{1.0}
    \caption{
      Theoretical synthetic spectra for \ofour\ with 
$T_{\rm eff}$=40kK, $\log L/L_{\odot}$=5.98, $v \sin i$=80 km\,s$^{-1}$
differing only in
      $\dot{M}$: 6$\times$10$^{-6}$yr$^{-1}$ (red, dotted),
      8.5$\times 10^{-6}$yr$^{-1}$ (green, dashed), and 11$\times 
10^{-6}$yr$^{-1}$ (blue, dot-dashed).
UVES observations are indicated as solid lines.
Notice the sensitivity of some 
features --
principally He\,{\sc ii} $\lambda$4686 and H$\alpha$ -- 
to mass-loss, whilst 
others (e.g. N\,{\sc iii} $\lambda\lambda$4634--41) are unaffected. 
     \label{mdot}
     } 
\end{figure}


\begin{figure}[h!]
    \epsscale{1.0}
    \caption{
     Comparison between optical (UVES) observations of 
     {\ofour} (O4\,Iaf$^+$) and synthetic spectroscopy
($v \sin i$=80 km\,s$^{-1}$)
     for a range of stellar temperatures, 
with mass-loss
rates adjusted to reproduce the observed H$\alpha$ emission profile.
    Models with processed (solid) and unprocessed (0.4$Z_{\odot}$,
dotted) CNO abundances are indicated in each panel. 
The blue optical lines of He\,{\sc i} $\lambda$4471, N\,{\sc iii} 
$\lambda\lambda$4634-41, N\,{\sc iv} $\lambda$4058, 
N\,{\sc v} $\lambda\lambda$4603-20 
are very sensitive to temperature and abundance.
    In addition to the usual helium and metal lines, our models include 
the newly identified S\,{\sc iv} $\lambda\lambda$4486--4504 lines
(Werner \& Rauch 2001).
     \label{hde269698_t}
     } 
\end{figure}

\begin{figure}[h!]
    \epsscale{1.0}
    \caption{
     Comparison between UV ({\hst}) observations of 
     {\ofour} (O4\,Iaf$^+$) and synthetic spectroscopy
($v \sin i$=80 km\,s$^{-1}$)
     for a range of stellar temperatures, with mass-loss rates
adjusted to reproduce the observed H$\alpha$ emission profile.
    Models with processed (solid) and unprocessed (0.4$Z_{\odot}$,
dotted) CNO abundances are indicated for each model. In contrast
with the  optical, UV P~Cygni lines are rather insensitive to temperature
and abundance variations. The greatest morphological differences 
over the temperature range covered are due to the weak iron 
features -- 
  Fe\,{\sc vi} $\lambda\lambda$1250--1350, 
  Fe\,{\sc v}  $\lambda\lambda$1350--1500, and 
  Fe\,{\sc iv} $\lambda\lambda$1550-1700.
For comparison purposes, the synthetic
model is corrected for Ly$\alpha$ at 1215.7\AA, 
with $\log (H_{2}$/cm$^{-2}$)=20.7.  
     \label{hde269698_t3}
     } 
\end{figure}

\begin{figure}
    \epsscale{1.0}
    \caption{
     Comparison between FUV (FUSE) observations of 
     {\ofour} (O4\,Iaf$^+$) and synthetic spectroscopy
($v \sin i$=80 km\,s$^{-1}$)
     for a range of stellar temperatures, with mass-loss rates
again adjusted to reproduce the observed H$\alpha$ emission profile.
    Models with processed (solid) and unprocessed (0.4$Z_{\odot}$,
dotted) CNO abundances are indicated for each model. The P~Cygni
profiles sampled in the FUSE spectral range, in contrast with
the {\hst} range, are very sensitive to temperature and abundances variations,
notably C\,{\sc iii} $\lambda$977, N\,{\sc iii} $\lambda\lambda$989-91, O\,{\sc vi}
$\lambda\lambda$1032-8, S\,{\sc iv} $\lambda\lambda$1062-72, P\,{\sc v} $\lambda\lambda$1118-28 and
C\,{\sc iii} $\lambda$1175. For comparison purposes, the
 synthetic model is corrected for the Lyman HI series 
(principally Ly$\beta$ at 1025.7\AA, and Ly$\gamma$ at 972.5\AA) 
with $\log (H_{2}$/cm$^{-2}$)=20.7. 
     \label{hde269698_t2}
     } 
\end{figure}

\begin{figure}[h!]
    \epsscale{1.0}
    \caption{
     Comparison between optical UVES line profiles of 
     {\ofour} (solid, O4\,Iaf$^+$) 
      and
     synthetic {\sc cmfgen} spectra (red, 
     dotted) for
     our final parameters, $T_{\rm eff}$=40kK, $\log (L/L_{\odot})$=5.98,
     $\log g$=3.6, $\dot{M}$=8.5$\times 10^{-6}$ $M_{\odot}$yr$^{-1}$, 
$v \sin i$=80 km\,s$^{-1}$, 
     $\beta$=1 and $v_{\infty}$=1750 km\,s$^{-1}$. 
      \label{hde269698_cmfgen}
     } 
\end{figure}

\begin{figure}[h!]
    \epsscale{1.0}
    \caption{
     Comparison between UV {\hst} and FUV {\fuse} observations  
     of {\ofour} (solid, O4\,Iaf$^+$) and synthetic {\sc cmfgen} spectra 
     (red, dotted) for
     our final parameters, $T_{\rm eff}$=40kK, $\log (L/L_{\odot})$=5.98,
     $\log g$=3.6, $\dot{M}$=8.5$\times 10^{-6}$ $M_{\odot}$yr$^{-1}$, 
$v \sin i$=80 km\,s$^{-1}$, 
     $\beta$=1 and $v_{\infty}$=1750 km\,s$^{-1}$. The synthetic
model is corrected for the Lyman HI series (principally Ly$\alpha$ at 
1215.7\AA, Ly$\beta$ at 1025.7\AA, and Ly$\gamma$ at 972.5\AA) according
to Herald et al. (2001) with $\log (H_{2}$/cm$^{-2}$)=20.7.  
      \label{hde269698_cmfgen3} 
      } 
\end{figure}



\begin{figure}[h!]
    \epsscale{1.0}
    \caption{
     Comparison between optical UVES line profiles of 
     {\osix} (solid, O6\,Iaf$^+$) and synthetic {\sc cmfgen} spectra (red, dotted) for
     our final parameters, $T_{\rm eff}$=33.5kK, $\log 
     (L/L_{\odot})$=5.86,
     $\log g$=3.2, $\dot{M}$=11$\times 10^{-6}$ $M_{\odot}$yr$^{-1}$, 
     $\beta$=1.3,
$v \sin i$=80 km\,s$^{-1}$, 
 and $v_{\infty}$=1520 km\,s$^{-1}$. 
     \label{hde270952_cmfgen}
     } 
\end{figure}


\begin{figure}[h!]
    \epsscale{1.0}
    \caption{
     Comparison between UV {\hst} and FUV {\fuse} observations of
     {\osix} (solid, O6\,Iaf$^+$) and synthetic {\sc cmfgen} spectra (red, 
     dotted) for
     our final parameters, $T_{\rm eff}$=33.5kK, $\log 
     (L/L_{\odot})$=5.86,
     $\log g$=3.2, $\dot{M}$=11$\times 10^{-6}$ $M_{\odot}$yr$^{-1}$, 
     $\beta$=1.3,
$v \sin i$=80 km\,s$^{-1}$, 
 and $v_{\infty}$=1520 km\,s$^{-1}$. The synthetic
model is corrected for the Lyman HI series (principally Ly$\alpha$ at 
1215.7\AA, Ly$\beta$ at 1025.7\AA, and Ly$\gamma$ at 972.5\AA) according
to Herald et al. (2001) with $\log (H_{2}$/cm$^{-2}$)=21.0. 
     \label{hde270952_cmfgen3}
     } 
\end{figure}

\begin{figure}[h!]
    \epsscale{1.0}
    \caption{
     Comparison between optical UVES line profiles of 
     {\oseven} (solid, O7\,Iaf$^+$) and synthetic {\sc cmfgen} spectra 
(red, dotted) for
     our final parameters, $T_{\rm eff}$=32kK, $\log 
(L/L_{\odot})$=5.85,
     $\log g$=3.1, $\dot{M}$=4.5$\times 10^{-6}$ $M_{\odot}$yr$^{-1}$, 
     $\beta$=1.65,
$v \sin i$=80 km\,s$^{-1}$, 
 and $v_{\infty}$=1330 km\,s$^{-1}$. 
     \label{av232_cmfgen}
     } 
\end{figure}

\begin{figure}[h!]
    \epsscale{1.0}
    \caption{
     Comparison between UV {\hst} and FUV {\fuse} observations of 
     {\oseven} (solid, O7\,Iaf$^+$) and synthetic {\sc cmfgen} spectra 
(red, dotted) for
     our final parameters, $T_{\rm eff}$=32kK, $\log 
(L/L_{\odot})$=5.85,
     $\log g$=3.1, $\dot{M}$=4.5$\times 10^{-6}$ $M_{\odot}$yr$^{-1}$, 
     $\beta$=1.65,
$v \sin i$=80 km\,s$^{-1}$, 
 and $v_{\infty}$=1330 km\,s$^{-1}$. The synthetic
model is corrected for the Lyman HI series (principally Ly$\alpha$ at 
1215.7\AA, Ly$\beta$ at 1025.7\AA, and Ly$\gamma$ at 972.5\AA) according
to Herald et al. (2001) with $\log (H_{2}$/cm$^{-2}$)=21.0. 
     \label{av232_cmfgen3}
     } 
\end{figure}

\begin{figure}[h!]
    \epsscale{1.0}
    \caption{
     Comparison between optical UVES line profiles of 
     {\onine} (solid, O9.7\,Ia$^+$) and synthetic {\sc cmfgen} spectra 
(red, dotted) for
     our final parameters, $T_{\rm eff}$=26kK, $\log 
(L/L_{\odot})$=5.82,
     $\log g$=2.8, $\dot{M}$=6.0$\times 10^{-6}$ $M_{\odot}$yr$^{-1}$, 
     $\beta$=1.75,
$v \sin i$=100 km\,s$^{-1}$, 
 and $v_{\infty}$=1000 km\,s$^{-1}$. 
     \label{sk66169_cmfgen}
     } 
\end{figure}

\begin{figure}[h!]
    \epsscale{1.0}
    \caption{
     Comparison between UV {\iue} and FUV {\fuse} observations of 
     {\onine} (solid, O9.7\,Ia$^+$) and synthetic {\sc cmfgen} spectra 
(red, dotted) for
     our final parameters, $T_{\rm eff}$=26kK, $\log 
(L/L_{\odot})$=5.82,
     $\log g$=2.7, $\dot{M}$=6.0$\times 10^{-6}$ $M_{\odot}$yr$^{-1}$, 
     $\beta$=1.75,
$v \sin i$=100 km\,s$^{-1}$, 
 and $v_{\infty}$=1000 km\,s$^{-1}$. The synthetic
model is corrected for the Lyman HI series (principally Ly$\alpha$ at 
1215.7\AA, Ly$\beta$ at 1025.7\AA, and Ly$\gamma$ at 972.5\AA) according
to Herald et al. (2001) with $\log (H_{2}$/cm$^{-2}$)=20.5.  
     \label{sk66169_cmfgen3}
     } 
\end{figure}

\begin{figure}[h!]
    \epsscale{1.0}
    \caption{
     Comparison between observed {\fuse} P\,{\sc v} 
$\lambda\lambda$1118--28
     profiles in {\oseven} (solid) 
and our standard model parameters  (upper panel,
red dotted), and cases for
which the wind is clumped with a volume filling factor of $f$=0.1
(central panel, red dotted) or P/H is reduced by a factor of four (lower 
panel, red dotted).
     \label{av232_pv}
     } 
\end{figure}

\clearpage

\begin{figure}[h!]
    \epsscale{1.0}
    \caption{
      Temperature calibrations for O supergiants from various sources
(red: B\"{o}hm-Vitense 1981; green: Schmidt-Kaler 1982; blue: Howarth \& 
Prinja 1989; black: Vacca
et al. 1996), universally based on plane-parallel hydrostatic results,
versus our spectroscopic determinations, allowing for the stellar wind
and line blanketing. Minor revisions to these calibrations at 
the earliest subtypes are needed due to the revisions of 
Walborn et al. (2002a). For example, HD\,93129A, one of two O3 supergiants 
included in the Vacca et al. calibration, has subsequently been 
re-classified to O2\,If*.
     \label{osuper}
     } 
\end{figure}

\begin{figure}[h!]
    \epsscale{1.0}
    \caption{
     Reduced wind momentum ($\dot{M} v_{\infty} R^{0.5}$, cgs units) versus 
     luminosity ($\log L/L_{\odot}$) for Galactic (circles), LMC 
(red, squares) and 
SMC (blue, triangles)
     O stars. Open symbols refer to data from Puls et al. (1996), whilst
     filled in symbols are from the present work (two in common -- arrows
     indicate revision in parameters). Also shown is the form of the 
     wind-momentum luminosity relationship (dotted line) for 
     Galactic O supergiants according to Kudritzki \& Puls (2000).
     \label{momentum}
     } 
\end{figure}

\end{document}